\documentclass[12pt]{article}
\usepackage{graphics,graphicx}
\usepackage{amssymb,epsfig,amsmath,euscript,array}
\usepackage{cite}
\usepackage{axodraw}
\usepackage{pstricks}
\usepackage{color}

\makeatletter
\@addtoreset{equation}{section}
\makeatother

\newcounter{multieqs}

\newcommand{\be}{\begin{equation}}
\newcommand{\ee}{\end{equation}}

\newcommand{\bm}[1]{\mbox{\boldmath $#1$}}

\newcommand{\kslash}{k \!\!\! / }

\newcommand{\lslash}{l \!\! / }
\newcommand{\Pslash}{P \!\!\!\! / }

\newcommand{\islash}{i \!\!\! / }
\newcommand{\jslash}{j \!\!\! / }
\newcommand{\aslash}{a \!\!\! / }
\newcommand{\bslash}{{b \hspace{-6pt} \slash} }

\newcommand{\onslash}{1 \!\!\! / }
\newcommand{\twslash}{2 \!\!\!/ }
\newcommand{\thslash}{3 \!\!\!/ }
\newcommand{\foslash}{4 \!\!\! / }
\newcommand{\fislash}{5 \!\!\! / }

\newcommand{\mslash}{m \!\!\! / }

\def\bd{\begin{document}}
\def\ed{\end{document}}
\def\nn{\nonumber}
\def\bea{\begin{eqnarray}}
\def\eea{\end{eqnarray}}

\def\ab{(ijab)}
\def\ba{(ijba)}
\def\ijab{{\tr}_{+}(\islash\, \jslash\, \aslash \, \bslash)}
\def\ijba{{\tr}_{+}(\islash\, \jslash\, \bslash \, \aslash)}
\def\ijaP{{\tr}_{+}(\islash\, \jslash\, \aslash \, \Pslash)}
\def\ijPLa{{\tr}_{+}(\islash\, \jslash\, \Pslash_L \, \aslash)}
\def\ijaPL{{\tr}_{+}(\islash\, \jslash\, \aslash \, \Pslash_L)}
\def\ijPLza{{\tr}_{+}(\islash\, \jslash\, \Pslash_{L;z} \, \aslash)}
\def\ijaPLz{{\tr}_{+}(\islash\, \jslash\, \aslash \, \Pslash_{L;z})}
\def\ijPa{{\tr}_{+}(\islash\, \jslash\, \Pslash \, \aslash)}
\def\iaPb{{\tr}_{+}(\islash\, \aslash\, \Pslash \, \bslash)}
\def\ibPa{{\tr}_{+}(\islash\, \bslash\, \Pslash \, \aslash)}
\def\ijPmu{{\tr}_{+}(\islash\, \jslash\, \Pslash \, \mu)}
\def\ibmuP{{\tr}_{+}(\islash\, \bslash\, \mu \, \Pslash)}
\def\ibmua{{\tr}_{+}(\islash\, \bslash\, \mu \, \aslash)}
\def\iamub{{\tr}_{+}(\islash\, \aslash\, \mu \, \bslash)}
\def\jaPb{{\tr}_{+}(\jslash\, \aslash\, \Pslash \, \bslash)}
\def\ijmuP{{\tr}_{+}(\islash\, \jslash\, \mu \, \Pslash)}
\def\ijmum{{\tr}_{+}(\islash\, \jslash\, \mu \, \mslash)}
\def\ijmmu{{\tr}_{+}(\islash\, \jslash\, \mslash \, \mu)}
\def\ijmP{{\tr}_{+}(\islash\, \jslash\, \mslash \, \Pslash)}
\def\iabP{{\tr}_{+}(\islash\, \aslash\, \bslash \, \Pslash)}
\def\ijbP{{\tr}_{+}(\islash\, \jslash\, \bslash \, \Pslash)}
\def\jbPa{{\tr}_{+}(\jslash\, \bslash\, \Pslash \, \aslash)}
\def\ijPb{{\tr}_{+}(\islash\, \jslash\, \Pslash \, \bslash)}
\def\jbmua{{\tr}_{+}(\jslash\, \bslash\, \mu \, \aslash)}

\def\loablt{ {\tr}_{+}(\lslash_1\, \aslash \, \bslash\, \lslash_2)}

\def\ijlolt{{\tr}_{+}(\islash\, \jslash\, \lslash_1 \, \lslash_2)}
\def\ijltlo{{\tr}_{+}(\islash\, \jslash\, \lslash_2 \, \lslash_1)}
\def\ibloa{{\tr}_{+}(\islash\, \bslash\, \lslash_1 \, \aslash)}
\def\jaltb{{\tr}_{+}(\jslash\, \aslash\, \lslash_2 \, \bslash)}
\def\ialtb{{\tr}_{+}(\islash\, \aslash\, \lslash_2 \, \bslash)}
\def\bltloa{{\tr}_{+}(\bslash\, \lslash_2\, \lslash_1 \, \aslash)}
\def\jbloa{{\tr}_{+}(\jslash\, \bslash\, \lslash_1 \, \aslash)}
\def\ibPb{{\tr}_{+}(\islash\, \bslash\, \Pslash \, \bslash)}
\def\ijltb{{\tr}_{+}(\islash\, \jslash\, \lslash_2 \, \bslash)}

\def\ijloa{{\tr}_{+}(\islash\, \jslash\,  \lslash_1 \, \aslash)}
\def\ijblt{{\tr}_{+}(\islash\, \jslash\,  \bslash \, \lslash_2)}

\def\jakb{{\tr}_{+}(\jslash\, \aslash\, \kslash \, \bslash)}
\def\iakb{{\tr}_{+}(\islash\, \aslash\, \kslash \, \bslash)}

\def\tofo{{\tr}_{+}(\onslash\, \thslash\, \twslash \, \foslash)}
\def\foto{{\tr}_{+}(\onslash\, \thslash\, \foslash \, \twslash)}
\def\tofi{{\tr}_{+}(\onslash\, \thslash\, \twslash \, \fislash)}
\def\fito{{\tr}_{+}(\onslash\, \thslash\, \fislash \, \twslash)}

\def\lrangle#1#2{\langle #1\,#2\rangle}

\def\Li{{$\rm Li}_2$}
\def\eps{\epsilon}
\def\epsuv{{\epsilon_{\rm \mbox{\tiny UV}}}}
\let\bm=\bibitem
\let\la=\label

\def\npb#1#2#3{Nucl. Phys. {\bf{B#1}} #3 (#2)}
\def\plb#1#2#3{Phys. Lett. {\bf{#1B}} #3 (#2)}
\def\prl#1#2#3{Phys. Rev. Lett. {\bf{#1}} #3 (#2)}
\def\prd#1#2#3{Phys. Rev. {D \bf{#1}} #3 (#2)}
\def\cmp#1#2#3{Comm. Math. Phys. {\bf{#1}} #3 (#2)}
\def\cqg#1#2#3{Class. Quantum Grav. {\bf{#1}} #3 (#2)}
\def\nppsa#1#2#3{Nucl. Phys. B (Proc. Suppl.) {\bf{#1A}}#3 (#2)}
\def\ap#1#2#3{Ann. of Phys. {\bf{#1}} #3 (#2)}
\def\ijmp#1#2#3{Int. J. Mod. Phys. {\bf{A#1}} #3 (#2)}
\def\rmp#1#2#3{Rev. Mod. Phys. {\bf{#1}} #3 (#2)}
\def\mpla#1#2#3{Mod. Phys. Lett. {\bf A#1} #3 (#2)}
\def\jhep#1#2#3{J. High Energy Phys. {\bf #1} #3 (#2)}
\def\atmp#1#2#3{Adv. Theor. Math. Phys. {\bf #1} #3 (#2)}
\newcommand{\EQ}[1]{\begin{equation} #1 \end{equation}}
\newcommand{\AL}[1]{\begin{subequations}\begin{align} #1 \end{align}\end{subequations}}
\newcommand{\SP}[1]{\begin{equation}\begin{split} #1 \end{split}\end{equation}}
\newcommand{\ALAT}[2]{\begin{subequations}\begin{alignat}{#1} #2 \end{alignat}
                        \end{subequations}}
\def\beqa{\begin{eqnarray}}
\def\eeqa{\end{eqnarray}}
\def\beq{\begin{equation}}
\def\eeq{\end{equation}}
\def\sst{\scriptscriptstyle}
\def\thetabar{\bar\theta}
\def\Tr{{\rm Tr}}
\def\one{\mbox{1 \kern-.59em {\rm l}}}
 \def\Nh{\hat{N}}

\newcommand{\half}{{\textstyle {1 \over 2}}}

\def\a{\alpha}      \def\da{{\dot\alpha}}
\def\b{\beta}       \def\db{{\dot\beta}}
\def\c{\gamma}  \def\G{\Gamma}  \def\cdt{\dot\gamma}
\def\d{\delta}  \def\D{\Delta}  \def\ddt{\dot\delta}
\def\e{\epsilon}        \def\vare{\varepsilon}
\def\f{\phi}    \def\F{\Phi}    \def\vvf{\f}
\def\h{\eta}
\def\k{\kappa}
\def\l{\lambda} \def\L{\Lambda}
\def\m{\mu} \def\n{\nu}
\def\o{\omega}
\def\p{\pi} \def\P{\Pi}
\def\r{\rho}
\def\s{\sigma}  \def\S{\Sigma}
\def\t{\tau}
\def\th{\theta} \def\Th{\Theta} \def\vth{\vartheta}
\def\X{\Xeta}
\def\z{\zeta}
\def\de{\partial}

\def\cA{{\cal A}} \def\cB{{\cal B}} \def\cC{{\cal C}}
\def\cD{{\cal D}} \def\cE{{\cal E}} \def\cF{{\cal F}}
\def\cG{{\cal G}} \def\cH{{\cal H}} \def\cI{{\cal I}}
\def\cJ{{\cal J}} \def\cK{{\cal K}} \def\cL{{\cal L}}
\def\cM{{\cal M}} \def\cN{{\cal N}} \def\cO{{\cal O}}
\def\cP{{\cal P}} \def\cQ{{\cal Q}} \def\cR{{\cal R}}
\def\cS{{\cal S}} \def\cT{{\cal T}} \def\cU{{\cal U}}
\def\cV{{\cal V}} \def\cW{{\cal W}} \def\cX{{\cal X}}
\def\cY{{\cal Y}} \def\cZ{{\cal Z}}

\def\ua{\underline{\alpha}}
\def\ub{\underline{\phantom{\alpha}}\!\!\!\beta}
\def\uc{\underline{\phantom{\alpha}}\!\!\!\gamma}
\def\um{\underline{\mu}}
\def\ud{\underline\delta}
\def\ue{\underline\epsilon}
\def\una{\underline a}\def\unA{\underline A}
\def\unb{\underline b}\def\unB{\underline B}
\def\unc{\underline c}\def\unC{\underline C}
\def\und{\underline d}\def\unD{\underline D}
\def\une{\underline e}\def\unE{\underline E}
\def\unf{\underline{\phantom{e}}\!\!\!\! f}\def\unF{\underline F}
\def\unm{\underline m}\def\unM{\underline M}
\def\unn{\underline n}\def\unN{\underline N}
\def\unp{\underline{\phantom{a}}\!\!\! p}\def\unP{\underline P}
\def\unq{\underline{\phantom{a}}\!\!\! q}
\def\unQ{\underline{\phantom{A}}\!\!\!\! Q}
\def\unH{\underline{H}}

\def\As {{A \hspace{-6.4pt} \slash}\;}
\def\bs {{b \hspace{-6.4pt} \slash}\;}
\def\Ds {{D \hspace{-6.4pt} \slash}\;}
\def\ds {{\del \hspace{-6.4pt} \slash}\;}
\def\ss {{\s \hspace{-6.4pt} \slash}\;}
\def\ks {{ k \hspace{-6.4pt} \slash}\;}
\def\ps {{p \hspace{-6.4pt} \slash}\;}
\def\pas {{{p_1} \hspace{-6.4pt} \slash}\;}
\def\pbs {{{p_2} \hspace{-6.4pt} \slash}\;}
\def\Ps {{P \hspace{-6.4pt} \slash}\;}
\def\Qs {{Q \hspace{-6.4pt} \slash}\;}

\def\Fh{\hat{F}}
\def\Vh{\hat{V}}
\def\Xh{\hat{X}}
\def\ah{\hat{a}}
\def\xh{\hat{x}}
\def\yh{\hat{y}}
\def\ph{\hat{p}}
\def\xih{\hat{\xi}}
\def\psit{\tilde{\psi}}
\def\Psit{\tilde{\Psi}}
\def\tht{\tilde{\th}}
\def\lt{\tilde{\lambda}}
\def\hl{\hat{\lambda}}
\def\hlt{\hat{\tilde{\lambda}}}
\def\llt{\tilde{l}}
\def\At{\tilde{A}}
\def\Qt{\tilde{Q}}
\def\Rt{\tilde{R}}
\def\Nt{\tilde{N}}

\def\at{\tilde{a}}
\def\st{\tilde{s}}
\def\ft{\tilde{f}}
\def\pt{\tilde{p}}
\def\qt{\tilde{q}}
\def\vt{\tilde{v}}
\def\nt{\tilde{n}}

\def\delb{\bar{\partial}}
\def\bz{\bar{z}}
\def\bD{\bar{D}}
\def\bB{\bar{B}}

\def\bk{{\bf k}}
\def\bl{{\bf l}}
\def\bp{{\bf p}}
\def\bq{{\bf q}}
\def\br{{\bf r}}
\def\bx{{\bf x}}
\def\by{{\bf y}}
\def\bR{{\bf R}}
\def\bV{{\bf V}}

\def\d{\delta}\def\D{\Delta}\def\ddt{\dot\delta}
\def\pa{\partial} \def\del{\partial}
\def\xx{\times}
\def\uno{\mbox{1 \kern-.59em {\rm l}}}
\def\trp{^{\top}}
\def\inv{^{-1}}
\def\dag{{^{\dagger}}}
\def\pr{^{\prime}}
\def\lan{\langle}
\def\ran{\rangle}
\def\rar{\rightarrow}
\def\lar{\leftarrow}
\def\lrar{\leftrightarrow}
\newcommand{\0}{\,\!}      
\def\one{1\!\!1\,\,}
\def\im{\imath}
\def\jm{\jmath}
\newcommand{\tr}{\mbox{tr}}
\newcommand{\slsh}[1]{/ \!\!\!\! #1}
\def\vac{|0\rangle}
\def\lvac{\langle 0|}
\def\hlf{\frac{1}{2}}
\def\ove#1{\frac{1}{#1}}
\def\Box{\square}
\def\ZZ{\mathbb{Z}}
\def\CC#1{({\bf #1})}
\def\bcomment#1{}
\def\bfhat#1{{\bf \hat{#1}}}
\def\VEV#1{\left\langle #1\right\rangle}
\newcommand{\ex}[1]{{\rm e}^{#1}} \def\ii{{\rm i}}
\def\rr{{\rm r}} \def\rs{{\rm s}}\def\rv{{\rm v}}
\def\ri{{\rm i}}\def\rj{{\rm j}}
\newcommand{\lrbrk}[1]{\left(#1\right)}
\newcommand{\sfrac}[2]{{\textstyle\frac{#1}{#2}}}

\def\Li{{\rm Li}_2}

\def\l{\langle}
\def\r{\rangle}

\font\mybb=msbm10 at 12pt
\def\bb#1{\hbox{\mybb#1}}

\font\myBB=msbm10 at 18pt
\def\BB#1{\hbox{\myBB#1}}

\begin{document}

\begin{flushright}
QMUL-PH-09-12
\end{flushright}

\vspace{20pt}

\begin{center}

{\Large \bf One-loop $\mathcal{N}=8$ supergravity coefficients    }
\\
\vspace{0.3cm}
{\Large \bf     from  $\mathcal{N}=4$ super Yang-Mills  }
\vspace{11pt}
\vspace{32pt}

{\mbox {\bf Panagiotis Katsaroumpas, Bill Spence  and Gabriele Travaglini}}%
\footnote{
{\sffamily \{\tt p.katsaroumpas, w.j.spence, g.travaglini\}@qmul.ac.uk }}

{\em Centre for Research in String Theory\\
Department of Physics\\
Queen Mary, University of London\\
Mile End Road, London, E1 4NS\\
United Kingdom
}

\vspace{30pt} {\bf Abstract}

\end{center}

\noindent
We use supersymmetric generalised unitarity to calculate  supercoefficients of box functions
in the expansion of scattering amplitudes in $\cN=8$ supergravity at one loop.
Recent advances have presented tree-level amplitudes in $\cN=8$ supergravity in terms
of sums of terms containing squares of colour-ordered Yang-Mills superamplitudes.
We develop the consequences of these results for the structure of one-loop supercoefficients, 
recasting them  as sums of squares of  $\cN=4$ Yang-Mills expressions with certain coefficients 
inherited from the tree-level superamplitudes.  
This provides new expressions for all one-loop box coefficients in $\cN=8$ supergravity, which we check against 
known results in a number of cases. 

\noindent

\setcounter{page}{0}
\thispagestyle{empty}
\newpage


\section{Introduction}
\setcounter{footnote}{0}

Recent advances have indicated that gravity scattering  amplitudes are much simpler than what one would infer from the 
Feynman diagram expansion, very much like in Yang-Mills theory.
In \cite{bbst,cs}, on-shell recursion relations were written down for graviton amplitudes at tree level,
and a remarkably benign ultraviolet behaviour of the scattering amplitudes under
certain large deformations along complex directions in momentum space  was observed.
This behaviour, not apparent from a simple analysis based on Feynman diagram considerations \cite{bbst,cs} similar to those discussed in \cite{bcfw} for Yang-Mills amplitudes, was later re-examined and explained in \cite{ben,ah1,cahk}.

At the quantum  level, unexpected cancellations occur in maximal supergravity
starting at one loop, which led to the conjecture \cite{Bern:1998sv, Bern:2005bb,BjerrumBohr:2005xx, BjerrumBohr:2006yw}
and later proof \cite{cahk,BjerrumBohr:2008vc} of  the  ``no-triangle  hypothesis".
According to this property, all one-loop amplitudes in $\cN=8$ supergravity can be written as sums of box functions
times rational coefficients, similarly to one-loop amplitudes in $\cN=4$ super Yang-Mills (SYM).
Interesting connections were established in \cite{bsgf3} and
\cite{BjerrumBohr:2008vc,Badger:2008rn} between one-loop cancellations, and the  large-$z$ behaviour observed in \cite{bbst,cs,ben,ah1,cahk}, as well as the
presence of summations over different orderings of the external particles typical of unordered theories such as gravity (and QED).
There is therefore growing  evidence of the remarkable similarities  between
the two maximally supersymmetric theories,
$\cN=4$ SYM and $\cN=8$ supergravity, culminating in the conjecture that the
$\cN=8$ theory could be ultraviolet finite, just like its non-gravitational  maximally supersymmetric counterpart.
This is supported both by  multi-loop perturbative calculations
\cite{bsgf1,bsgf2,bsgf3,Bern:2009kd},
and  string theory and M-theory considerations \cite{Green:2006gt,Green:2006yu,chalmers,berko}.

In  a recent paper \cite{Elvang:2007sg}, Elvang and Freedman  were able to recast $n$-graviton
MHV  amplitudes at tree level in a suggestive form in terms of   sums of squares of $n$-gluon MHV amplitudes.
An analytical proof for all $n$ of the agreement of their expression to that  for the infinite sequence of MHV amplitudes conjectured from recursion relations in \cite{bbst} was also presented, as well as numerical checks showing agreement with the  Berends-Giele-Kuijf formula \cite{bgk}. A direct proof of the  formula of \cite{Elvang:2007sg} was later given in \cite{Spradlin:2008bu}.

In a related development at tree level, the authors of \cite{Drummond:2009ge} used supersymmetric recursion relations
\cite{bhtrec, cahk} of the BCF type \cite{bcfrec,bcfw}, and the explicit solution found in the  $\cN=4$ case in
\cite{Drummond:2008cr},
to recast amplitudes in   \(\cN=8\) supergravity  in a new simplified form which involves sums of $\cN=4$ amplitudes. 
Specifically,  according to \cite{Drummond:2009ge}  a generic $\cN=8$ superamplitude
can be written as
\beq
\label{prima}
\cM (1, \ldots , n)\ = \ \sum_{\cP(2, \ldots , n-1)} M(1, \ldots , n)
\ ,
\eeq
where the ordered subamplitudes $M(1, \ldots , n)$ are \cite{Drummond:2009ge}:
\beq
\label{dsvwformula}
M(1, \ldots , n) \ = \ [ A^\mathrm{MHV}(1 , \ldots , n) ]^2 \, \sum_{\alpha} [ R_{\alpha}(\lambda_i, \tilde{\lambda}_i, \eta_i)]^2 \,
G_{\alpha} (\lambda_i, \tilde{\lambda}_i )
\ .
\eeq
Here $A^\mathrm{MHV}(1 , \ldots , n)$ is the MHV superamplitude in $\cN=4$ SYM \cite{Nair}, and
$R_{\alpha}$ are certain dual superconformal invariant  quantities   \cite{Drummond:2008cr},
extending those introduced in \cite{dhks,dhksgen} for the  next-to-MHV (NMHV)  superamplitudes.
$G_{\alpha} $ are certain gravity  ``dressing factors", which are independent of the superspace variables
$\eta_i$ associated to each particle $i$ in the amplitude.
Finally, the sum in \eqref{prima} is over all  permutations of the labels $(2, \ldots , n-1)$. 
The fact that the sum over permutations in \eqref{prima} does not contain two of the $n$ scattered particles will be important in what  follows. 

Turning to loop amplitudes, it has been shown recently in \cite{dhksgen}  that four-dimensional
generalised unitarity \cite{bcfgen} may be efficiently applied to calculate the supercoefficients
of one-loop superamplitudes  in  \(\cN=4\) SYM. One of the advantages of the use of superamplitudes is that it
makes it particularly efficient to  perform the sums over internal helicities
\cite{Georgiou:2004wu,bst,Massimo,Elvang:2008na,Elvang:2008vz,dhksgen,cahk,Bern:2009xq}, 
which are converted into fermionic  integrals. 
Furthermore, according to the no-triangle  property of maximal supergravity
\cite{Bern:1998sv , Bern:2005bb,BjerrumBohr:2005xx, BjerrumBohr:2006yw,BjerrumBohr:2008vc,cahk}, 
one-loop amplitudes  in the $\cN=8$ theory are expressed in terms of box functions only, 
therefore the coefficients of one-loop amplitudes can be calculated 
by using  quadruple cuts. It is therefore natural to  investigate how 
the new expressions for generic tree-level \(\cN=8\) supergravity amplitudes  found in  \cite{Drummond:2009ge}
can be used together with supersymmetric quadruple cuts  \cite{dhksgen} in order 
to derive new formulae for one-loop amplitudes in \(\cN=8\) supergravity. This will be the main goal of this paper. 

One interesting consequence of the structure of \eqref{prima} for the results 
we derive for the one-loop box supercoefficients 
is that when the expressions for tree-level amplitudes are inserted into quadruple cuts, 
they give rise to new general formulae for  the supercoefficients
that are  written  as sums of squares of the result of the corresponding 
\(\cN=4\) SYM  calculation (apart from the four-mass case, this will be 
the square of an $\cN=4$ coefficient), multiplied by certain dressing factors.   
The one-loop supercoefficients therefore inherit 
the intriguing structure of tree-level amplitudes  exhibited by  \eqref{prima} and \eqref{dsvwformula}.

Specifically, we will calculate supercoefficients for MHV, NMHV and $\mathrm{N^2MHV}$ superamplitudes, 
and we will show in a number of cases how these new expressions
match known formulae. 
In particular, we will show how our results agree with  the  expressions for the infinite sequence of MHV amplitudes obtained in \cite{Bern:1998sv} using unitarity, with the five-point NMHV amplitude \cite{Bern:1998sv}, and with  
the six-point graviton NMHV amplitudes coefficients 
derived in \cite{Bern:2005bb,BjerrumBohr:2005xx}. 
In the MHV case, we propose a correspondence between the ``half-soft" functions introduced  in \cite{Bern:1998sv} 
and particular sums of dressing factors, which we  check numerically up to 12 external legs.  
In \cite{Bern:1998sv,Bern:2005bb,BjerrumBohr:2005xx},
the tree-level amplitudes entering the cut had been generated using  KLT relations
\cite{Kawai:1985xq}; in our approach, we will instead use the solution of the supersymmetric recursion relation
given by \eqref{prima} and \eqref{dsvwformula}. Our results support the conjecture that all one-loop amplitude
coefficients in \(\cN=8\) supergravity may be written in terms of
\(\cN=4\) Yang-Mills expressions times known dressing factors.

The rest of the  paper is organised as follows. In the next section we will briefly review some background material needed in order to describe amplitudes in maximally supersymmetric theories, and quadruple cuts.
In Section 3 we will study MHV superamplitudes at one loop, deriving a straightforward  general
expression for the supercoefficients in the $n$-point case. We propose a conjecture which enables an immediate
correspondence to be made with the known general formula for these amplitudes, and test this
explicitly for $2me$ coefficients with up to $n=22$ external legs.
Section 4 turns to consider NMHV amplitudes. We derive general expressions for
the $3m$ and $2me$ box coefficients, and the related $2mh$ and $1m$ coefficients. Similarly to the SYM case considered in \cite{dhksgen}, all the supercoefficients can be written in terms of the $3m$ coefficients, which we are able to recast as sums of squares of the corresponding SYM $3m$ coefficients, times certain bosonic dressing factors.
In Section 5 we
study explicit examples, starting with the five-point NMHV case, which provides a simple toy model
for studying structures at higher points, and we then discuss the six-point NMHV case. 
In Section 6 we describe how this approach applies in general to $\mathrm{N^{\it p}MHV}$ amplitude coefficients.
We conclude with some discussion of further work.

{\bf Note added:} 
After this paper was completed, we became aware of    \cite{Hall:2009xg}, which appears to overlap with our paper. 

\section{Background}

In the supersymmetric formalism of   \cite{Nair},   one associates to each particle in the $\cN=8$ theory
the usual commuting spinors
$\lambda_\a$, $\lt_{\da}$ (in terms of which the momentum of the $i^\mathrm{th}$
particle is $p_{\a \da}^{i} = \lambda_{\a}^{i} \lt_{\da}^{i}$), as well
as anticommuting variables $\eta^{A}_{i}$, where $A=1,\ldots,8$ is an $SU(8)$ index.
The supersymmetric amplitude can then be expanded in powers of the
$\cN=8$ superspace coordinates $\eta_{A}^{i}$ for the different particles,
and each term of this expansion corresponds to a particular scattering
amplitude in $\cN=8$ supergravity. In particular, a term containing $m_i$ powers of
$\eta_{A}^{i}$ corresponds to a scattering process where the
$i^\mathrm{th}$  particle has helicity $h_i = 2 - m_i/2$.

Generalising the discussion of \cite{dhksgen} to \(\cN=8\) supergravity, we
write a generic $n$-point superamplitude with $n>3$ as
\begin{equation}
  \cM_n(\lambda, \tilde{\lambda}, \eta)
\  = \ i (2\pi)^4 \delta^{(4)}(p) \, \delta^{(16)}(q )
  \ \mathcal{P}_n(\lambda, \tilde{\lambda}, \eta),
\end{equation}
where the function \(\mathcal{P}_n\) has the form
\begin{equation}
  \mathcal{P}_n
  =\mathcal{P}_n^{(0)}+\mathcal{P}_n^{(8)}
   +\mathcal{P}_n^{(16)}+\ldots+\mathcal{P}_n^{(8n-32)},
\end{equation}
where \(\mathcal{P}_n^{(8k)}(\lambda, \tilde{\lambda}, \eta)\) is an \(SU(8)\) invariant
homogenous polynomial in the \(\eta\)'s of degree \(8k\). Furthermore
$p := \sum_{i=1}^n  \lambda_i \lt_i$ is the total momentum of the particles, and
$q^A_\alpha := \sum_{i=1}^n q^A_{\alpha; i} $ is the sum of the supermomenta
$q^A_{\alpha; i} := \eta^A_i  \lambda_{\alpha; i} $ of each particle $i$.
The fermionic delta function \(\delta^{(16)}(q_{\alpha}^{A})\) raises the degree of each term
to \(8k+16\). Each \(\eta_i^A\) carries helicity of
\(1/2\),   therefore the term \(\delta^{(16)}(q_{\alpha}^{A})\mathcal{P}_n^{(8k)}(\lambda, \tilde{\lambda}, \eta)\)
has total helicity of \(2k+8\), giving the N$^k$MHV amplitude.

The three-point supergravity amplitudes are given by \cite{cahk}
\beqa
\label{eq:threepointMHV}
{\mathcal{M}}^{\rm MHV}_3 (1,2,3) & = &   \big[A^{{\rm MHV}}_3 (1,2,3)\big]^2 \ = \
\frac{\delta^{(16)}(\sum_{i=1}^n \eta_i\lambda_i)}{(\l 12 \r \l23\r \l31\r)^2}\ ,
\\
\label{eq:threepointantiMHV}
{\mathcal{M}}^{\overline{\rm MHV}}_3 (1,2,3) & = &  \big[A^{\overline{\rm MHV}}_3 (1,2,3)\big]^2 \ = \
\frac{ \delta^{(8)}(
\eta_1 [ 23] + \eta_2 [31] + \eta_3 [12] ) }
{ ([12][23][31])^2 }\, ,
\eeqa
and are obtained by simply squaring the corresponding MHV \cite{Nair} and anti-MHV \cite{bhtrec,cahk} superamplitudes in
$\cN=4$  SYM. Notice in \eqref{eq:threepointantiMHV} the presence of an unusual   fermionic delta function.
It is also easy to show that \eqref{eq:threepointantiMHV} is invariant under all supersymmetries.

One-loop amplitudes can be expanded in a known basis of scalar integrals which,
in maximally supersymmetric theories, contains only
box functions. This was shown in  \cite{bddk} for the SYM case, and is the content of the no-triangle property
mentioned earlier.
We will therefore write a generic one-loop superamplitude
in maximal supergravity as
\begin{equation}
  {\mathcal{M}}^{\mathrm{1-loop}}_n \ =  \    
  \sum\mathcal{C}_n(P_1,P_2,P_3,P_4)\,  \cI(P_1,P_2,P_3,P_4)\ ,
\end{equation}
where we are summing over all distinct scalar box functions $\cI(P_1,P_2,P_3,P_4)$  with external momenta $P_1, \ldots , P_4$
\cite{Bern:1993kr},  and  $\mathcal{C}_n(P_1,P_2,P_3,P_4)$ are the supercoefficients of the expansion.

Using generalised unitary in four dimensions, the one-loop supercoefficients of
maximal supergravity amplitudes can be expressed in terms of products of four tree-level amplitudes as
\beqa
\label{supercut}
  &&\mathcal{C}(P_1,P_2,P_3,P_4)
 \,  =\,
  \frac{1}{2}
  \sum_{\cS_{\pm}}\int\!\prod_{i=1}^4 d^{8}\eta_{l_i}
  \nonumber\\
  &&\qquad \times\
  \cM(-l_1,P_1,l_2)\, \cM(-l_2,P_2,l_3)\,
\cM(-l_3,P_3,l_4)\, \cM(-l_4,P_4,l_1)\, ,
\eeqa
where we are averaging over the two solutions \(\cS_{\pm}\) to  the cut
conditions \(l_{i}^{2}=0\), $i=1, \ldots , 4$,  which impose that
all the internal propagators are on shell \cite{bcfgen}, and
supermomentum conservation delta functions for each of the four amplitudes entering \eqref{supercut}  are understood.
The integration is performed over the four Grassmann variables $\eta_{l_i}$, $i=1, \ldots , 4$ associated with  the internal cut legs.

In the next sections we will describe how \eqref{supercut} can be applied  to obtain supercoefficients for the MHV and NMHV amplitudes in maximal supergravity.

\section{One-loop MHV superamplitudes  }

In \cite{dhksgen}, the  one-loop MHV superamplitude in $\cN =4$ SYM was derived using a supersymmetric extension of the  quadruple unitarity cuts of \cite{bcfgen}. It turns out that many of the details of the calculation
presented in \cite{dhksgen}  carry over directly to the supergravity case, and we will follow closely the notation of these authors in order
to simplify the comparison.

\begin{figure}[ht]
\begin{center}
\scalebox{0.70}{\includegraphics{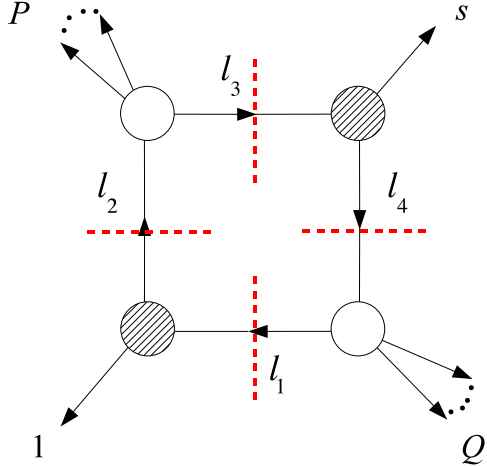}}
\end{center}
\caption{\it  A quadruple cut diagram determining the supercoefficient $\mathcal{C}^{\cN=4}(1,P,s,Q)$ in the MHV superamplitudes in
$\cN=4$ SYM and $\cN=8$ supergravity.  The two three-point amplitudes
have the anti-MHV helicity configurations, and the remaining two amplitudes are MHV's. }
\label{MHVquad}
\end{figure}

We begin by giving the expression derived in \cite{dhksgen}
for the supercoefficient of the generic diagram contributing to the MHV superamplitude, 
drawn in Figure \ref{MHVquad},
\beqa\label{eq:MHVcoeff}
\mathcal{C}^{\cN=4}(1,P,s,Q) &=&  \frac12 \sum_{\mathcal{S}_\pm}\int \prod_{i=1}^4 d^4\eta_{l_i}  \\ \nonumber
&&\!\!\!\!\!\!\!\times\ \frac{\delta^{(4)}(\eta_1 [l_2 l_1] + \eta_{l_2}[l_1 1] +\eta_{l_1}[1
    l_2])}{[1 l_2][l_2 l_1][l_1 1]}
\frac{\delta^{(8)}(\lambda_{l_2}\eta_{l_2} + \sum_2^{s-1} \lambda_i \eta_i
  - \lambda_{l_3} \eta_{l_3})}{\l l_2 2\r\cdots\l s-1\,\,l_3\r \l l_3
  l_2\r }  \\ \nonumber
&&\!\!\!\!\!\!\!\times\ \frac{\delta^{(4)}(\eta_{l_3}[s l_4] + \eta_s[l_4 l_3] + \eta_{l_4}[l_3
    s])}{[l_3 s][s l_4][l_4 l_3]}\frac{\delta^{(8)}(\lambda_{l_4}
  \eta_{l_4} + \sum_{s+1}^n \lambda_i \eta_i -
  \lambda_{l_1}\eta_{l_1})}{\l l_4\,\,s+1\r \cdots\l n l_1\r \l l_1 l_4\r },
\eeqa
where the sum goes over the two solutions to the cut equations.
The four terms in \eqref{eq:MHVcoeff} come from the product of the four tree-level amplitudes in the
diagram,
\beq
{A}_3^{\overline{\rm MHV}} (-l_1,1,l_2)\ A(-l_2,2,\dots,s-1,l_3)\ {A}_3^{\overline{\rm MHV}}(-l_3,s,l_4)\
A^\mathrm{MHV}(-l_4,s+1,\cdots,n,l_1),
\eeq
where $A^\mathrm{MHV}$ is the MHV superamplitude \cite{Nair},
\beq
\label{MHVNair}
A^\mathrm{MHV}_{n} (1, \ldots , n )  = {\delta^{(8)} \big( \sum_{i=1}^{n} \eta_{i} \lambda_{i } \big)
 \over N(1, 2, \ldots , n)
}
\ ,
\eeq
and we have defined
\beq\label{eq:spinorproduct}
N(1,2, \ldots ,n) := \l 12\r \l 23\r \cdots \l n1 \r
\ .
\eeq

As shown in \cite{dhksgen}, the result of evaluating \eqref{eq:MHVcoeff}  is
\beq
\label{eq:MHVresult}
\mathcal{C}^{\cN=4}(1,P,s,Q) \ = \
{1\over 2} (P^2Q^2 - s t )\,
\frac{\delta^{(8)}(\sum_{i=1}^n\eta_i\lambda_i)}{N(1,2,\ldots ,n)}\, .
\eeq
Incidentally, we notice that in order to arrive at \eqref{eq:MHVresult} it is not necessary to know
the explicit solutions to the cut equations, but only that the holomorphic spinors at the
three-point anti-MHV corners are proportional, i.e.~$\lambda_{l_1} \propto \lambda_{l_2} \propto \lambda_1$, $\lambda_{l_3} \propto \lambda_{l_4} \propto \lambda_s$.
We also notice that for $n>4$ only one of the two cut solutions contributes to the cut diagram of  Figure \ref{MHVquad}. 
This leads to the factor of $1/2$ on the right-hand side of \eqref{eq:MHVresult}.

Let us now consider the same quadruple-cut diagram but in the $\cN=8$ supergravity case.
The supercoefficient is given by the following integral,
\beqa
\nonumber
\mathcal{C}^{\cN=8}(1,P,s,Q)&\!\! =\!\! &\frac{1}{2}\sum_{\mathcal{S}_\pm}\int \prod_{i=1}^4 d^8\eta_{l_i}
 \ {\cM}_3^{\overline{\mathrm{MHV}}}(-l_1,1,l_2)\
 \cM^\mathrm{MHV}(-l_2,2,\dots,s-1,l_3)\
 \\
 \label{eq:MHVgrav}
 &&{\cM}_3^{\overline{\mathrm{MHV}}}(-l_3,s,l_4)\
\cM^\mathrm{MHV}(-l_4,s+1,\dots,n,l_1),
\eeqa
where we sum over the same two solutions of the cut equations as in the Yang-Mills case, but now integrate over the eight superspace variables $\eta^A$, $A=1, \ldots , 8$, and insert the tree-level MHV and anti-MHV supergravity amplitudes $\cM^\mathrm{MHV}$ and ${\cM}^{\overline{\mathrm{MHV}}}$.

The simplification in calculating \eqref{eq:MHVgrav} comes when we
utilise the result of Elvang and Freedman \cite{Elvang:2007sg},
who have given the following expression for the $n$-point ($n>3)$ tree-level MHV supergravity amplitudes
in terms of the Yang-Mills MHV tree amplitudes and certain \lq\lq dressing factors\rq\rq\ $G^{\rm MHV}$:
\beq
\label{eq:ElvangFreedmanMHV}
{\mathcal{M}}^{\rm MHV}_n = \sum_{{\mathcal{P}}(2,\ldots,n-1)}
[A^{\rm MHV}(1,\ldots,n)]^2
G^{\rm MHV}(1,\ldots,n)\ ,
\eeq
where we sum over permutations ${\mathcal{P}}(2,\ldots,n-1)$ of the elements $(2,\dots,n-1)$,
and the dressing factors are given by
\beq
\label{eq:dressingfunctions}
G^{\rm MHV}(1,\ldots,n) =
x_{13}^2 \prod_{s=2}^{n-3}
\frac{\langle s | x_{s,s+2} x_{s+2,n} |n\rangle}{\l sn\r }\ .
\eeq
We have used in the above the form of the result as given in \cite{Drummond:2009ge}.
Note that the dressing factors are independent of the
superspace variables $\eta_i$.

Now we may insert these expressions into the formula \eqref{eq:MHVgrav} to find the
box coefficients $\mathcal{C}^{\cN=8}(1,P,s,Q)$. This results in a sum of products of squares
of Yang-Mills superamplitudes, times dressing factors. The key point here is that
since the dressing factors are independent of the superspace variables, we may
follow exactly the manipulations of \cite{dhksgen} in order to carry out the superspace integrations, and this will yield exactly the square of the Yang-Mills result \eqref{eq:MHVresult}. We follow the conventions of 
\cite{Drummond:2009ge} with regard to squaring delta functions, in particular  it is understood that
\beq
\bigg(\delta^{(8)}\big(\sum_{i=1}^n \eta_i\lambda_i\big)\bigg)^2 =
\delta^{(16)}\big(\sum_{i=1}^n \eta_i\lambda_i\big)
\ ,
\eeq
where there are the four Yang-Mills $\eta$ variables on the left-hand side of this expression
 and the eight supergravity ones on the right-hand side.

Hence the result of the superspace integrals in \eqref{eq:MHVgrav} is
\beqa\label{eq:MHVgravfinal}
&&\mathcal{C}^{\cN=8}(1,P,s,Q)= \frac{1}{2}(P^2Q^2-st)^2 \sum_{\mathcal{S}_\pm} \sum_{{\mathcal{P}}(2,\ldots,s-1)} \sum_{{\mathcal{P}}(s+1,\ldots,n)}  \\  \nonumber
&&\qquad\qquad \times \ \frac{G^{\rm MHV}(-l_2,2,\dots,s-1,l_3)  G^{\rm MHV}(-l_4,s+1,\dots,n,l_1)}{(N(1,\dots,n))^2}\ ,
\eeqa
where the dressing factors $G^\mathrm{MHV}$ are given by \eqref{eq:dressingfunctions}, and
the first summation involves inserting the explicit solutions to the quadruple cut conditions.
This solution is very easy to determine for the two-mass easy box case.
In the specific cut in Figure \ref{MHVquad}, where the  two three-point superamplitudes
have the anti-MHV helicity configuration,  there is only one solution for the  cut loop momenta, 
which has the form 
\beq
\label{cutmom}
l_1 \, = \, \lambda_1 \tilde\lambda_{l_1}\ , \qquad
l_2 \, = \, \lambda_1 \tilde\lambda_{l_2}\ , \qquad
l_3 \, = \, \lambda_s \tilde\lambda_{l_3}\ , \qquad
l_4 \, = \, \lambda_s \tilde\lambda_{l_4}\ , \qquad
\eeq
and we wish to determine $\tilde\lambda_{l_1}, \ldots , \tilde\lambda_{l_4} $.
This is accomplished  by imposing momentum conservation at the four corners of the cut diagram.
The result is
\beq
\label{cutmom2}
\tilde\lambda_{l_1} \, = \,  - { \lan s|Q \over \lan s 1 \ran} \ , \qquad
\tilde\lambda_{l_2} \, = \,  { \lan s|P\over \lan s 1 \ran} \ , \qquad
\tilde\lambda_{l_3} \, = \,  - { \lan 1|P \over \lan  1s \ran} \ , \qquad
\tilde\lambda_{l_4} \, = \,   { \lan 1|Q \over \lan  1 s\ran} \ , \qquad
\eeq
from which the cut momenta $l_1, \ldots , l_4$ are then obtained using \eqref{cutmom},
\beq
\label{cutmom3}
l_1 \, = \, - |1\r  { \lan s|Q \over \lan s 1 \ran}\ , \qquad
l_2 \, = \,  |1\r  { \lan s|P\over \lan s 1 \ran}\ , \qquad
l_3 \, = \,-  |s\r { \lan 1|P \over \lan  1s \ran}\ , \qquad
l_4 \, = \,  |s\r  { \lan 1|Q \over \lan  1 s\ran} \ . \qquad
\eeq
Given that only one solution to the cut contributes, one can drop the sum over ${\mathcal{S}_\pm}$ in
\eqref{eq:MHVgravfinal}, and a factor of $1/2$ is left over.

Taking this into account, we can instantly recast \eqref{eq:MHVgravfinal} as 
\beq
\mathcal{C}^{\cN=8}(1,P,s,Q)=
\sum_{{\mathcal{P}}(P)} \sum_{{\mathcal{P}}(Q)}
\Big[\mathcal{C}^{\cN=4}(1, P ,s,  Q)\Big]^2
\
2\, G^{\rm MHV}(-l_2,P,l_3)  G^{\rm MHV}(-l_4,Q,l_1) ,
\label{eq:MHVgravfinal3}
\eeq
where  $P$ and $Q$ here denote the sets $P=\{2,\dots,s-1\}$ and $Q=\{s+1,\dots n\}$.
The $\cN=4$ supercoefficient $\mathcal{C}^{\cN=4}(1, P ,s, Q)$ is given in \eqref{eq:MHVresult}, and 
the loop momenta are evaluated on the solution provided by \eqref{cutmom3}.
The expression \eqref{eq:MHVgravfinal3} gives a new form of the one-loop integral coefficients in the
supergravity MHV amplitudes for any number of external legs.

Next we would like to compare our result  \eqref{eq:MHVgravfinal3} to previously known expressions for the
MHV coefficients. In \cite{Bern:1998sv} the infinite sequence of  graviton MHV amplitudes  was presented.
The result of that paper for the two-mass easy coefficients is%
\footnote{A factor of $(-1)^n$  in the result of  \cite{Bern:1998sv} for the $2me$ coefficients
 can be attributed to different conventions.} 
\beq\label{eq:BDPRresult}
\mathcal{C}^{\cN=8}(1,P,s,Q) = {1 \over 2} \, \big(P^2Q^2-st\big)^2 h(1,\{P\},s)\ h(s,\{Q\},1)\ . 
\eeq
The first three
``half-soft" functions $h(a, M, b) $ are given by
\beqa\label{eq:hfunctions}
h(a,\{1\},b) &=& {1\over \l a1\r ^2 \l 1b\r ^2} \,, \\ \nonumber
h(a,\{1,2\},b) &=&
{ [12] \over
  \l 12\r \l a1 \r \l 1b\r \l a2\r \l 2b \r} \,, \\ \nonumber
h(a,\{1,2,3\},b) &=&
{ [12][23]\over
  \l 12\r \l 23\r  \l a1\r \l 1b\r \l a3\r \l 3b\r  }
\, +\,  { [23][31]\over
  \l 23\r \l 31\r  \l a2\r \l 2b\r \l a1\r \l 1b\r  }
\\ \nonumber
  &+&
  { [31][12]\over
  \l 31\r \l 12\r  \l a3\r \l 3b\r \l a2\r \l 2b\r  }
\,.
\eeqa
A recursive form for the $h$ functions is also given in \cite{Bern:1998sv}
as well as the following explicit formula:
\beqa
\nonumber
h(a,\{1,2,\ldots,n\},b)& =& \frac{[12]}{ \l 12\r }
 { \l a| K_{1,2} | 3]   \l a| K_{1,3}| 4]
   \cdots \l a | K_{1,n-1} | n]
  \over \l 23\r \l 34\r \cdots \l n-1 n\r
  \, \l a1\r
  \l a 2\r  \cdots \l a n \r
  \, \l 1b\r \l nb\r }\\
&+& \mathcal{P}(2,3,\ldots,n) \ ,
\label{eq:hfunctionsgeneral}
\eeqa
where $K_{i,j} = k_i + k_{i+1} + \cdots + k_j$.

In order to show that the two expressions \eqref{eq:MHVgravfinal3} and \eqref{eq:BDPRresult} for the one-loop MHV amplitude coefficients are equivalent, we consider a massive tree sub-amplitude in the loop diagram under consideration, for
example that containing the set of momenta $Q=\{s+1,\dots,n\}$, with internal
loop momenta $l_4, l_1$.
We now make the following conjecture relating the $h$ functions in \eqref{eq:BDPRresult} to the dressing factors $G$ of \eqref{eq:MHVgravfinal}:
\beq\label{eq:conjecture}
\sum_{\mathcal{P}(s+1,\dots,n)}\frac{ G^\mathrm{MHV}(-l_4,Q,l_1)}{\big(\l s s+1\r\l s+1 s+2\r \cdots\l n-1 n\r\l n1\r\big)^2}
= h(s,\{ Q\},1)\ ,
\eeq
where it is assumed that a solution to the cut loop momentum constraints is
inserted in the left-hand side of this equation.
If this relation is true, it follows directly that
our formula \eqref{eq:MHVgravfinal3} is identical  to \eqref{eq:BDPRresult}.
Let us first see how the equality \eqref{eq:conjecture} works in some simple cases.

For the case where $Q$ is a single momentum, the result \eqref{eq:conjecture} is immediate
since $G(a,b,c)=1$ and $h(a,\{b\},c) =1 / (\l ab\r \l bc\r)^2$.

\begin{figure}[ht]
\begin{center}
\scalebox{0.70}{\includegraphics{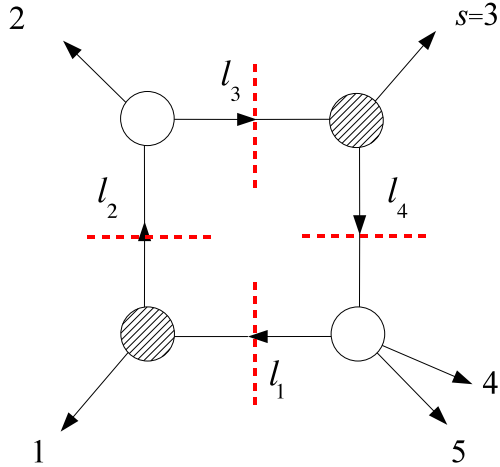}}
\end{center}
\caption{\it  The quadruple cut diagram considered for the derivation of {\rm \eqref{eq:thingy}}.
 }
\label{fig2}
\end{figure}

The next check we perform is for the case when $Q$ contains two momenta, where we  suppose that the labels of
the amplitude  are $(-l_4,4,5,l_1)$, with the neighbouring external legs being labeled $3$ and $1$, as in Figure \ref{fig2}.
Then we wish to show that
\beq\label{eq:thingy}
\sum_{\mathcal{P}(4,5)} \frac{ G^\mathrm{MHV}(-l_4,4,5,l_1)}{(\l 34\r \l 45\r \l 51\r)^2}
\, = \, h(3,\{4,5\},1).\
\eeq
The loop variable solution is in this case
\beq
\label{cutmomfive}
l_1\,  = \, \frac{\l 34\r  |1\r |4] + \l 35\r |1\r |5]}{\l 13\r} \ ,
\eeq
which follows from \eqref{cutmom3} with $Q = p_4 + p_5$. We also notice that the four-point dressing factor  is given by the same expression in \eqref{eq:dressingfunctions} but without the product, i.e. $G^\mathrm{MHV}(1,2,3,4) = x_{13}^2$.
Taking this into account, inserting \eqref{cutmomfive}  into the left-hand side of \eqref{eq:thingy}
and using standard identities, we arrive at
\beq
\label{stide}
\sum_{\mathcal{P}(4,5)} \frac{ G^\mathrm{MHV}(-l_4,4,5,l_1)}{(\l 34\r \l 45\r \l 51\r)^2}\, = \,
\frac{[45]}{\l 45\r} \frac{1}{ \l 34\r \l 41 \r \l 35\r \l 51\r }
\ ,
\eeq
which is precisely $h(3,\{4,5\},1)$. Note that we did not use any properties
of the momenta at the other massive tree amplitude  in any diagram containing this one.

For the next case,
let us suppose that the labels of its legs are $(l_4,4,5,6,l_1)$,
with the neighbouring external legs being labeled $3$ and $1$ again.
Then we wish to show that
\beq\label{eq:thingy2}
\sum_{\mathcal{P}(4,5,6)} \frac{ G^\mathrm{MHV}(-l_4,4,5,6,l_1)}{(\l 34\r \l 45\r \l 56\r\l 61\r)^2}
\, = \, h(3,\{4,5,6\},1)\ .
\eeq
The loop variable solution which we will use in this case follows again from \eqref{cutmom3} with
$Q = p_3 + p_4 + p_5$,
\beq
l_1 = \frac{1}{\l 13\r} \Big( \l 34 \r |1\r |4] + \l 35\r |1\r| 5] + \l 36 \r |1\r |6]\Big)\ .
\eeq
Inserting this into the left-hand side of \eqref{eq:thingy2} one finds
\beq
\sum_{\mathcal{P}(4,5,6)} \frac{ G^\mathrm{MHV}(-l_4,4,5,6,l_1)}{(\l 34\r \l 45\r \l 56\r\l 61\r)^2}\, = \,
\frac{1}{\l 13 \r} \sum_{\mathcal{P}(4,5,6)}
 \frac{[56]( [45]\l 15 \r + [46]\l 16 \r)}{\l 14\r \l 34\r \l 45 \r \l 16\r \l 56\r^2 }\ .
\eeq
Adding the terms from the permutations $(456)$ and $(465)$ it is straightforward
to obtain the first term of $h(3,\{4,5,6\},1)$ as given
using the last formula in \eqref{eq:hfunctions}; the cyclically rotated
terms are obtained in the same way.

\begin{figure}[ht]
\begin{center}
\scalebox{0.70}{\includegraphics{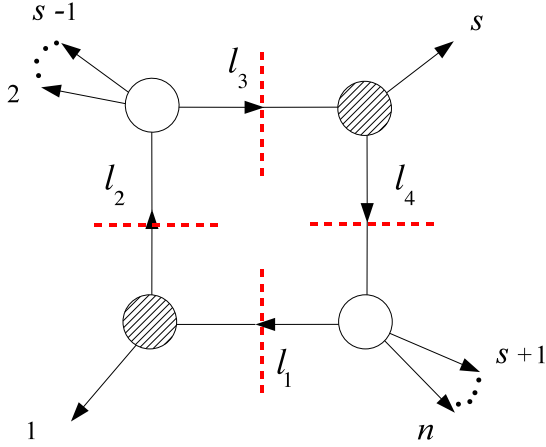}}
\end{center}
\caption{\it  The quadruple cut diagram determining the two-mass easy supercoefficient
$\mathcal{C}^{\cN=8}(1,P,s,Q)$ in the MHV amplitudes in $\cN=8$ supergravity where
$P:= \sum_{l=2}^{s-1}p_l$,  $Q :=\sum_{l=s+1}^{n}p_l$. The black blobs denote anti-MHV three-point amplitudes, the remaining amplitudes have the MHV helicity configuration.}
\label{2meMHVcoefffig}
\end{figure}

To study the conjecture \eqref{eq:conjecture} in general, we consider the quadruple cut diagram in
Figure \ref{2meMHVcoefffig}.
Again only one solution to the cut equations contributes; we will need explicit expressions for the cut momenta $l_1$ and $l_4$, which are given by
\beq
l_1=\frac{1}{\l 1 s \r}
    \sum_{i=s+1}^{n}
    \l si \r |1\r| i]\ ,\qquad
l_4=\frac{1}{\l s1 \r}
    \sum_{i=s+1}^{n}
    \l i 1 \r |s\r| i] \ .
\eeq
Consider first the dressing factor from the MHV amplitude $\cM^\mathrm{MHV}(-l_4, s+1, \ldots , n, l_1)$.
This is given by
\beq
G^\mathrm{MHV}(-l_4,s+1,\ldots,n,l_1)
\, =\, s_{-l_4,s+1}
 \prod_{r=s+1}^{n-2}
 \frac{\l r |(r+1)\, \sum_{r+2}^{n} i |l_1 \r}
      {\l r l_1\r}
      \ ,
\eeq
where
\beq
s_{-l_4,s+1}\, = \, \frac{1}{\l 1s \r}
          \sum_{i=s+1}^{n} \l i 1\r \l s\: s+1\r [s+1\: i]
\ .
\eeq
Inserting the solution for the cut loop momenta into $G^\mathrm{MHV}(-l_4,s+1,\ldots,n,l_1)$,
and denoting
 the corresponding quantity $G'(s,\{s+1,\ldots,n\},1)$, one finds
\beq
G'(s,\{s+1,\ldots,n\},1)
\, =  \, \frac{1}{\l 1s \r}
 \left(\sum_{i=s+1}^{n}  \l i 1\r \l s\: s+1\r [s+1\: i] \right)
 \prod_{r=s+1}^{n-2}
 \frac{\l r |(r+1)\, \sum_{r+2}^{n} i | 1\r}
      {\l r 1\r}\ .
\eeq
We have checked numerically that
\beq
\sum_{\mathcal{P}(s+1,\ldots,n)}
\frac{G'(s,\{s+1,\ldots,n\},1)}
     {\Big(\langle s (s+1)\rangle\langle (s+1) (s+2)\rangle
      \cdots \langle n 1 \rangle\Big)^2}
\, = \, h(s,\{s+1,\ldots,n\},1)
\ ,
\eeq
for up to 12 legs, i.e. for $n$ up to $s+10$.
Note that an identical  argument applies to the other  massive corner  with momentum $P:= \sum_{l=2}^{s-1} p_l$
in the $2me$ box diagram. Therefore, this numerical check shows that the
two expressions \eqref{eq:MHVgravfinal3} and \eqref{eq:BDPRresult} for the $2me$ coefficients are equivalent
for up to $22$ external legs, whereas for the $1m$ diagrams, the equivalence is up to $13$ legs.

With the loop solutions inserted, our expression for the MHV amplitudes
in supergravity is then given by
\beq
\mathcal{C}^{\cN=8}(1,P,s,Q)=
  \sum_{{\mathcal{P}}(P)} \sum_{{\mathcal{P}}(Q)}
   \left(\mathcal{C}^{\cN=4}(1,P,s,Q)\right)^2
   G'(1,P,s)\;
   G'(s,Q,1)\ .
\eeq
These results indicate that generalised unitarity works for the MHV superamplitudes at one loop, and utilising suitable expressions for the
tree amplitudes in this process, one derives expressions for the supergravity coefficients as sums of squares of
SYM coefficients times known dressing factors, as given in the equation above.

We now move on to consider NMHV superamplitudes.


\section{Next-to-MHV supergravity amplitudes}

For next-to-MHV amplitudes in $\cN=8~$ supergravity, the three-mass and two-mass hard box functions also appear, in addition to the two-mass easy  and one-mass ones.
The relevant quadruple cut diagrams are the same as those appearing in \cite{dhksgen} in the $\cN=4$ SYM
case. In the following, we will give general expressions for the different
box coefficients.

\subsection{Three-mass and two-mass hard coefficients}

\begin{figure}[ht]
\begin{center}
\scalebox{0.70}{\includegraphics{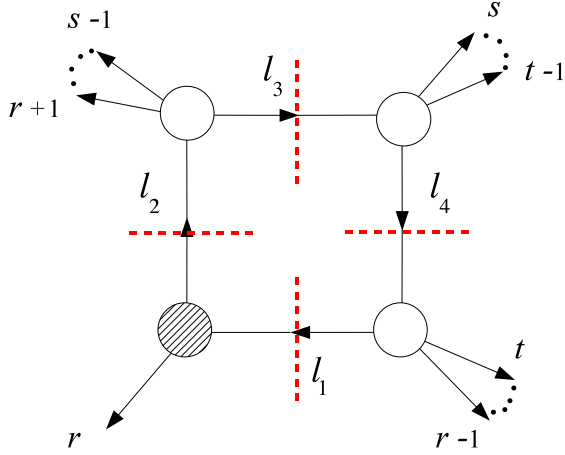}}
\end{center}
\caption{\it  The quadruple cut diagram determining the three-mass supercoefficient $\mathcal{C}^{\cN=8}(r,P,Q,R)$ in the NMHV amplitudes in $\cN=8$ supergravity.  There is a single  three-point amplitude participating in the cut, with the anti-MHV helicity configuration.    The remaining three superamplitudes are MHV's. We  also define
$P:= \sum_{l=r+1}^{s-1}p_l$,  $Q :=\sum_{l=s}^{t-1}p_l$, and $R:= \sum_{l=t}^{r-1}p_l$. }
\label{3mqc}
\end{figure}

We begin by considering three-mass coefficients. In this case, there is one quadruple cut diagram containing three MHV amplitudes and one anti-MHV, with each MHV amplitude containing more than three legs in general. The relevant quadruple cut diagram is represented in Figure \ref{3mqc} and yields
the expression
\beqa\label{eq:3mstart}
&&\!\!\!\!\!\!\!\!\!\!\mathcal{C}^{\cN=8}_{\rm 3m}(r,P,Q,R)= \frac12 \sum_{\mathcal{S}_\pm}\int \prod_{i=1}^4 d^8\eta_{l_i}
\ {\cM}_3^{\overline{\rm MHV}} (-l_1,r,l_2)\ \cM^\mathrm{MHV}(-l_2,r+1,\dots,s-1,l_3) \nonumber\\
 &&\qquad\quad \qquad\quad \times\  \cM^\mathrm{MHV}(-l_3,s,\dots,t-1,l_4)\
\cM^\mathrm{MHV}(-l_4,t,\cdots,r-1,l_1)\ ,
\eeqa
where the three-point anti-MHV amplitude is given in \eqref{eq:threepointantiMHV}. The MHV superamplitude
may be written in terms of squares of Yang-Mills amplitudes times
dressing factors using \eqref{eq:ElvangFreedmanMHV} and \eqref{eq:dressingfunctions}.
What will be important in what follows is that in the sum over permutations in \eqref{eq:ElvangFreedmanMHV}
there are always two missing legs. In applying this formula to write down explicitly the MHV superamplitudes entering the cut diagram in \eqref{eq:3mstart}, we will arrange these two missing legs to be precisely the loop legs.

Since the dressing factors are independent of the superspace variables $\eta$, the fermionic integrations in \eqref{eq:3mstart} can then be done similarly to those for the SYM case in
 \cite{dhksgen}.
Only one of the two solutions to the cut equations contributes to the maximal cut diagram  in Figure \ref{3mqc},
hence we can drop the sum over ${\mathcal{S}_\pm}$ in \eqref{eq:3mstart}, which then becomes
\beqa\label{eq:3mend}
&&\mathcal{C}^{\cN=8}_{\rm 3m}(r,P,Q,R)=
  \sum_{{\mathcal{P}}(P)} \sum_{{\mathcal{P}}(Q)}\sum_{{\mathcal{P}}(R)}
   \left(\mathcal{C}^{\cN=4}_{\rm 3m}(r,P,Q,R)\right)^2
   \nonumber\\
   &&\qquad\quad
   \times\ 2 \ G^{\rm MHV}(-l_2,P,l_3)\;
   G^{\rm MHV}(-l_3,Q,l_4)\;
   G^{\rm MHV}(-l_4,R,l_1)\ ,
\eeqa
where
\beq\label{eq:3mNequal4}
\mathcal{C}^{\cN=4}_{\rm 3m}(r,P,Q,R) = \frac{\delta^{(8)}\big(\sum_{i=1}^n (\eta_i\lambda_i)\big) R_{r;st}   }{ \prod_j\l j j+1\r }
\,
\Delta_{r,r+1,s,t}
\, ,
\eeq
is the corresponding Yang-Mills supercoefficient, calculated in  \cite{dhks}.
The dual superconformal invariants $R_{r; st}$ are given by
\cite{dhksgen,Drummond:2008cr}
\be\label{Rfunction}
R_{r;st} =  \frac{\l s-1 s\r \l t-1 t\r \  \delta^{(4)}\bigl(\Xi_{r;st}\bigr)}{x_{st}^2\l r|x_{rt}x_{ts}|s-1\r \l r|x_{rt}x_{ts}|s\r
\l r|x_{rs}x_{st}|t-1\r \l r|x_{rs}x_{st}|t\r }\, ,
\ee
where
\beq
\label{Xifunction}
\Xi_{r; st} \, := \,
\l r | \left[ x_{rs} x_{st} \sum_{k=t}^{r-1} |k\r \eta_k+ x_{rt} x_{ts} \sum_{k=s}^{r-1} | k \r \eta_k
\right]
\ ,
\eeq
and $x_{ab} := \sum_{l=a}^{b-1} p_l$. Finally,
\beq
\Delta_{r,r+1,s,t}\ = \ {1\over 2} \big( x_{rs}^2 x_{r+1 t}^2 - x_{rt}^2 x_{r+1 s}^2 \big)
\ .
\eeq
We have thus managed to express each three-mass coefficient  as a sum of squares of SYM coefficients, weighted with
bosonic dressing factors and summed over the appropriate permutations.

The product of three tree-level dressing factors in \eqref{eq:3mend} can in principle be further
simplified by inserting  the explicit solution to the cut expression.
The generic solution (when the four corners are massive) has been worked out in \cite{bcfgen}.
One can however find rather simple expressions in terms of spinor variables when at least one of the four amplitudes
participating in the quadruple cut is a three-point amplitude.
For the three-mass configuration, the quadruple cut solutions have been presented in \cite{lanceetal,kasperphd} in a compact form. For the specific case in Figure \ref{3mqc}, where the three-point amplitude is anti-MHV, the solution is  \cite{lanceetal,kasperphd}
\beqa
l_1 &=&{|r\r \l r| PQR|\over \l r | PR |r \r}\ , \qquad 
l_2 \ = \ {|r\r \l r| RQP|\over \l r | PR |r \r}\ ,
\\ \nonumber
l_3 &=& {|QR\, r\r \l r P|\over \l r | PR |r \r}\ , \qquad
\, l_4 \ = \ {|QP\, r\r \l r R|\over \l r | PR |r \r}\ ,
\eeqa
whereas the dressing factors are given by  \eqref{eq:dressingfunctions}, which in this case gives
\beqa
G^\mathrm{MHV}(-l_2, \{ P \}, l_3 ) &=&
s_{-l_2 r+1} \prod_{k=r+1}^{s-3} {\l k | x_{k, k+2} \, x_{k+2, l_3} | l_3 \r \over \l k l_3 \r}
\ , 
\\ \nonumber
G^\mathrm{MHV}(-l_3, \{ Q \}, l_4 ) &=&
s_{-l_3 s} \prod_{k=s}^{t-3} {\l k | x_{k, k+2} \, x_{k+2, l_4} | l_4 \r \over \l k l_4 \r}
\ ,
\\ \nonumber
G^\mathrm{MHV}(-l_4, \{ R \}, l_1 ) &=&
s_{-l_4 t} \prod_{k=s}^{r-3} {\l k | x_{k, k+2} \, x_{k+2, l_1} | l_1 \r \over \l k l_1 \r}
\ .
\eeqa

\begin{figure}[ht]
\begin{center}
\scalebox{0.70}{\includegraphics{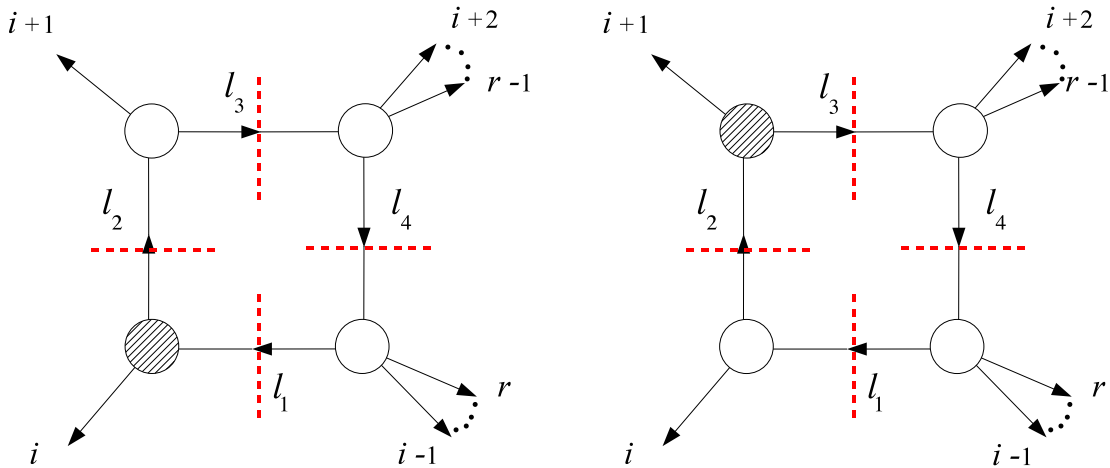}}
\end{center}
\caption{\it  The two quadruple cut diagrams determining the two-mass hard supercoefficient $\mathcal{C}^{\cN=8}(i,j,P,Q)$ in the NMHV amplitudes in $\cN=8$ supergravity.  Three-point amplitudes depicted in (black) white
have the (anti-)MHV helicity configuration.  The remaining two amplitudes are MHV's. }
\label{2mhcoeff}
\end{figure}

We now turn to the two-mass hard coefficients. There are two quadruple cut diagrams contributing here. These  are shown in Figure \ref{2mhcoeff}, where the two adjacent three-point amplitudes
are MHV and anti-MHV (or vice versa).
Similarly to the $\cN=4$ case discussed in \cite{dhksgen}, these two diagrams can be regarded as special cases of
the three-mass diagrams in  Figure \ref{3mqc}.
The result for the first diagram is simply given by $\mathcal{C}^{\cN=8}_{\rm 3m} (i,  i+1, P, Q)$, whereas for the second one has
$\mathcal{C}^{\cN=8}_{\rm 3m} ( i+1,  i ,  Q, P)$,
where $P:= \{p_{i+2}, \ldots , p_{ r-1}\} $ and $Q:= \{p_{r}, \ldots , p_{ i-1}\} $. The three-mass coefficients are defined in Figure
\ref{3mqc}.
The two-mass hard coefficients are then equal to
\beq\label{eq:2mhend}
\mathcal{C}^{\cN=8}_{\rm 2mh}(i,i+1,P,Q)\ = \
\mathcal{C}^{\cN=8}_{\rm 3m}(i,i+1,P,Q)  \, +\,
\mathcal{C}^{\cN=8}_{\rm 3m}(i+1, i , Q, P)
\ .
\eeq
We will present in Section \ref{6ptNMHV} some numerical checks of  \eqref{eq:2mhend} for the case of six-point NMHV superamplitudes, finding agreement with the results of \cite{Bern:2005bb,BjerrumBohr:2005xx}.

\subsection{Two-mass-easy and one-mass coefficients}

\begin{figure}[ht]
\begin{center}
\scalebox{0.70}{\includegraphics{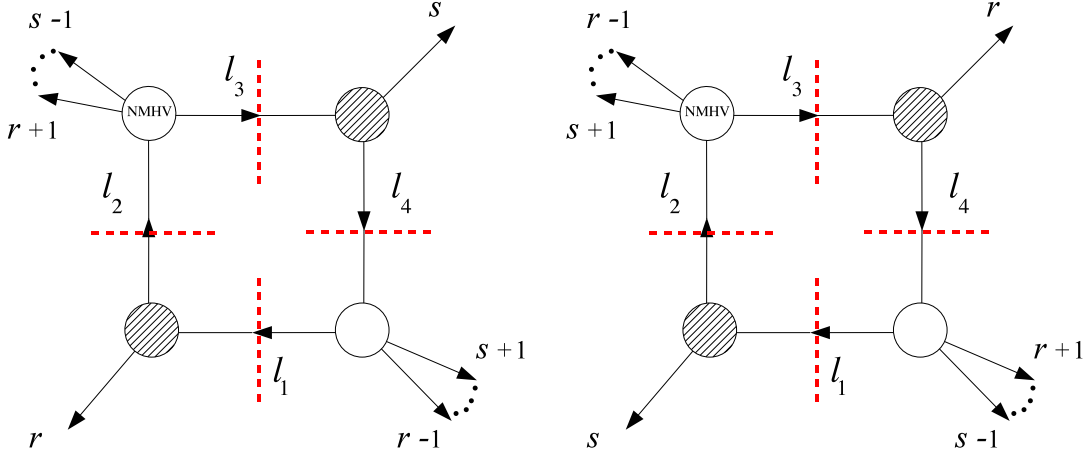}}
\end{center}
\caption{\it  The two quadruple cut diagrams determining the two-mass easy  supercoefficient $\mathcal{C}^{\cN=8}(r,P,s,Q)$ in the NMHV amplitudes in $\cN=8$ supergravity.  Three-point amplitudes have the anti-MHV helicity configuration, whereas the
white amplitudes are  MHV. We  also define
$P:= \sum_{l=r+1}^{s-1}p_l$, and $Q :=\sum_{l=s+1}^{r-1}p_l$. }
\label{2mecoefffig}
\end{figure}

We now move on to consider the two-mass easy coefficients, and as a particular case of these, the  one-mass coefficients.
In the two-mass easy case there are two diagrams, as in the SYM case considered in  \cite{dhksgen},
related to each other by a simple exchange of labels. Each cut diagram has two anti-MHV amplitudes, one NMHV amplitude and one MHV amplitude, see Figure \ref{2mecoefffig}.%
\footnote{An additional quadruple cut can actually be constructed by replacing one of the two three-point
anti-MHV amplitude with a three-point MHV one, and compensating this by replacing further the NMHV amplitude by an MHV one.
It can easily be shown \cite{dhksgen} that this particular quadruple cut would lead to constraints on the external kinematics, and hence can be ignored.}

Consider the first diagram. The result from the quadruple cut is
\beqa\label{eq:2mestart}
&&\!\!\!\!\!\!\!\!\!\!\!\!\!\!\!
\left.
\vphantom{\sqrt{2}}
\mathcal{C}^{\cN=8}_{\rm 2me}(r,P,s,Q)
\right|_{1}
= \frac12 \sum_{\mathcal{S}_\pm}\int \prod_{i=1}^4 d^8\eta_{l_i}
{\cM}_3^{\overline{\rm MHV}} (-l_1,r,l_2)\
\cM^{\rm NMHV}(-l_2, {r+1} , \ldots , {s-1} ,l_3) \nonumber\\
 &&\qquad\quad \qquad\quad \times\  {\cM}_3^{\overline{\rm MHV}}(-l_3,s,l_4)\
\cM^{\rm MHV}(-l_4,{s+1} , \ldots , {r-1} ,l_1)\ .
\eeqa

\begin{figure}[ht]
\begin{center}
\scalebox{0.70}{\includegraphics{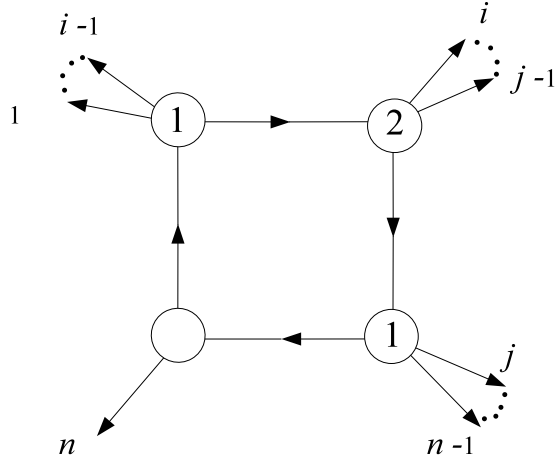}}
\end{center}
\caption{\it  Diagrammatic representation of the superconformal invariant $R_{n; ij}$. The numbers in the blobs indicate the minimum number of legs which need to be attached to that blob. From this Figure, it is easy to  see that $\Xi_{n; ij}$ in \eqref{Xifunction} does not depend either on
$\eta_n$ and $\eta_1$. }
\label{Rfigure}
\end{figure}

Here we may use the expression for NMHV tree amplitudes given in
\cite{Drummond:2009ge}, namely
 \beq
 \label{boo}
 \cM^{\rm NMHV} (1, \ldots , n) =  \sum_{\cP(2, \ldots , n-1)} M^{\rm NMHV}(1, \ldots , n)\ ,
 \eeq
where the   ordered subamplitude $M^{\rm NMHV}(1, \ldots , n)$  is
\begin{equation}
\label{NMHV}
M^{\rm NMHV}(1,\ldots,n) = [A^{\rm MHV}(1,\ldots,n)]^2
\sum_{i = 2}^{n-3} \sum_{j=i+2}^{n - 1} R_{n;ij}^2\; G^{\rm NMHV}_{n;ij}\ .
\end{equation}
The explicit expressions for the dressing factors  $G^{\rm NMHV}_{n;ij}$
are given in \cite{Drummond:2009ge}, and
$R_{n; ij}$ are the  dual superconformal invariants given in \eqref{Rfunction};
a useful diagrammatic representation of these quantities
was suggested in \cite{dhksgen}, and is reproduced for convenience here in Figure \ref{Rfigure}.

The very important property of  $R_{n; ij}$, which follows immediately from its definition \eqref{Rfunction}
and from Figure \ref{Rfigure},  is that it does not depend either on  $\eta_n$ and $\eta_1$.
This observation simplifies drastically our  calculation, and
is eventually responsible for why we will be able to write the $\cN=8$ supercoefficients as sums of squares of SYM supercoefficients.
Specifically, in writing explicitly the  tree-level NMHV superamplitude $\cM^{\rm NMHV}(-l_2, p_{r+1} , \ldots , p_{s-1} ,l_3)$ in
\eqref{eq:2mestart} using \eqref{boo} and \eqref{NMHV},   we will pick the loop legs $-l_2$ and $l_3$ to be $1$ and $n$ appearing in the latter formulae.  Two important consequences of this are that, firstly,
the sum over permutations in \eqref{boo} will not involve the cut-loop legs $-l_2$ and $l_3$; and, secondly, that
the supermomenta $\eta_{l_2}\lambda_{l_2}$ and $\eta_{l_3}\lambda_{l_3}$ of the cut legs will appear only through the overall supermomentum conservation delta functions.    Therefore,  the fermionic integrations over
$\eta_{l_2}$ and $\eta_{l_3}$  in \eqref{supercut}
will  proceed as in the case of the  supergravity  MHV superamplitude discussed previously.%
\footnote{The same property was observed in the $\cN=4$ calculation of \cite{dhks}. It is quite remarkable that this property continues to hold in maximal supergravity.}

We now proceed with the calculation.
Inserting  \eqref{boo} and  \eqref{NMHV}  into \eqref{eq:2mestart}, as well as the expressions for the three-point anti-MHV amplitude \eqref{eq:threepointantiMHV}  and for the MHV amplitude in \eqref{eq:ElvangFreedmanMHV}, we find
\beqa\label{eq:NMHV2meExpr1}
  &&\!\!\!\!\!\!\!
  \mathcal{C}^{\mathcal{N}=8\text{ (NMHV)}}_{\rm 2me}(r,P,s,Q)
  =\frac{1}{2}\sum_{S_{\pm}}\int \prod_{i=1}^{4}d^8\eta_{l_i}
  \left[A_3^{\overline{\text{MHV}}}(-l_1,r,l_2)\right]^2\,
  \left[A_3^{\overline{\text{MHV}}}(-l_3,s,l_4)\right]^2
  \nonumber\\
  &&
  \qquad
  \times
  \sum_{\mathcal{P}(P)}
  \left(
  \left[A^{\text{MHV}}(-l_2,\{P\},l_3)\right]^2
      \sum_{i=r+1}^{s-3}\sum_{j=i+2}^{s-1}
      (R_{l_3;ij})^2 G^{\text{NMHV}}_{l_3;ij}
  \right)
  \nonumber\\
  &&
  \qquad
  \times
  \sum_{\mathcal{P}(Q)}
   \left[A^{\text{MHV}}(-l_4,\{Q\},l_1)\right]^2
   G^{\text{MHV}}(-l_4,\{Q\},l_1)
  \ +   \ r \leftrightarrow s
  \ ,
\eeqa
where the first ($r \leftrightarrow s$) term in \eqref{eq:NMHV2meExpr1} corresponds to the
cut diagram on the left (right) of Figure \ref{2mecoefffig}.
By $\{P\}$, $\{Q\}$, we mean the (ordered) sets  of momenta
$(p_{r+1}, \ldots , p_{s-1})$, and $( p_{s+1}, \ldots, p_{r-1})$.

Next we observe that only one of the two cut solutions contributes, namely the solution in \eqref{cutmom3}.
We can then recast \eqref{eq:NMHV2meExpr1} as
\beqa\label{eq:NMHV2meExpr2}
  &&\!\!\!\!\!\!\!
  \mathcal{C}^{\mathcal{N}=8\text{ (NMHV)}}_{\rm 2me}(r,P,s,Q)
  =2\sum_{\mathcal{P}(\{P\})}\sum_{\mathcal{P}(\{Q\})}
  \left[\mathcal{C}^{\mathcal{N}=4\text{ (MHV)}}_{\rm 2me}(r,P,s,Q)\right]^2
  \\
  &&
  \qquad\qquad
  \times
  \left(
      \sum_{i=r+1}^{s-3}\sum_{j=i+2}^{s-1}
      (R_{l_3;ij})^2 G^{\text{NMHV}}_{l_3;ij}
  \right)
  G^{\text{MHV}}(-l_4,Q,l_1)
  +r \leftrightarrow s
  \ .
\nonumber
\eeqa
The dual superconformal invariant $R$-function appearing in \eqref{eq:NMHV2meExpr2} is given by \eqref{Rfunction}.
Explicitly,
\beq
\label{Rabove}
 R_{l_3,ij}\ = \
 \frac{\langle i-1 i\rangle \langle j-1 j\rangle
       \delta^{(4)}(\Xi_{l_3;ij})}
      {x_{ij}^2
       \langle l_3| x_{r+1,i} x_{ij}|j \rangle
       \langle l_3| x_{r+1,i} x_{ij}|j-1 \rangle
       \langle l_3| x_{r+1,j} x_{ji}|i \rangle
       \langle l_3| x_{r+1,j} x_{ji}|i-1 \rangle}\ ,
\eeq
where
\beq
  \Xi_{l_3;ij}\ = \
  -\langle l_3 |
  \left(
    x_{r+1,i}x_{ij}\sum_{m=j}^{s-1}|m\rangle \eta_m
   +x_{r+1,j}x_{ji}\sum_{m=i}^{s-1}|m\rangle \eta_m
  \right)
\ .
\eeq
A few comments are  in order here.

Firstly, we need to insert the cut solutions into the previous expressions.
These are obtained from  \eqref{cutmom3} by just replacing  $1\rightarrow r$.
Furthermore, when  the minimum value of $i$, i.e.~$i=r+1$ is attained in the sum appearing in \eqref{eq:NMHV2meExpr2}, the corresponding spinor for $i-1$ is actually
$|i-1 \rangle\equiv |-l_2 \rangle$, since the $R$-function comes from
the NMHV amplitude with legs
$(-l_2, r+1 , \ldots , s-1, l_3)$. However, the expression for $R$ in
\eqref{Rabove} is invariant under rescalings of  $|i-1\rangle$.
Hence, since    $| l_2\r \propto |r \r$ because of the cut condition, we conclude that
we can set  $|i-1\rangle  \to  |r\rangle$
when the minimum value in the sum over $i$ in \eqref{eq:NMHV2meExpr2} is attained.

Furthermore, we notice that     $| l_3\r \propto |s \r$ because of the cut condition. By expanding
the fermionic delta function $ \delta^{(4)} (\Xi_{l_3;ij})$ we see that this will contribute   four powers of
  $\langle l_3|$; inspecting \eqref{Rabove}, we conclude that
$R_{l_3,ij}$ will eventually be invariant under rescalings of  $\langle l_3|$ as well. We can then replace
 $\langle l_3 |\to \langle s |$ inside the expression for $R_{l_3; i j }$ or, equivalently,
 $R_{l_3; i j }\to R_{s; i j }$ and  $\Xi_{l_3; i j }\to \Xi_{s; i j }$, so that the explicit loop solutions
are not present in these quantities.

 Taking into account the previous remarks, we arrive at
 \beqa\label{eq:NMHV2meExpr222}
  &&\!\!\!\!\!\!\!
  \mathcal{C}^{\mathcal{N}=8\text{ (NMHV)}}_{\rm 2me}(r,P,s,Q)
  =2\sum_{\mathcal{P}(P)}\sum_{\mathcal{P}(Q)}
  \left[\mathcal{C}^{\mathcal{N}=4\text{ (MHV)}}_{\rm 2me}(r,P,s,Q)\right]^2
  \nonumber\\
  &&
  \qquad\qquad
  \times
  \left(
      \sum_{i=r+1}^{s-3}\sum_{j=i+2}^{s-1}
      (R_{s;ij})^2 G^{\text{NMHV}}_{l_3;ij}
  \right)
  G^{\text{MHV}}(-l_4,Q,l_1)
  \, +\, r \leftrightarrow s
  \ .
\eeqa
 The general expressions for the MHV dressing factors are  given in
 \eqref{eq:dressingfunctions}, from which one can obtain $G^{\text{MHV}}(-l_4,Q,l_1)$.

Finally we consider one-mass coefficients. As explained in \cite{dhksgen} in the Yang-Mills case,
the two relevant diagrams are special cases of other diagrams. In the first of them,
the three three-point corners are MHV-$\overline{\text{MHV}}$-MHV and the fourth corner is MHV, which
is a special case of the NMHV $2me$ coefficient. In the second diagram, the three
three-point corners are $\overline{\text{MHV}}$-MHV-$\overline{\text{MHV}}$ and the fourth
corner is NMHV, which is a special case of the NMHV three-mass coefficient.
Therefore, one finds
\beq\label{eq:1mend}
\mathcal{C}^{\cN=8}_{\rm 1m}(s+2,P,s,s+1) =
\mathcal{C}^{\cN=8}_{\rm 2me}(s+2,P,s,s+1)  +
\mathcal{C}^{\cN=8}_{\rm 3m}(s+1,s+2,P,s)
\ .
\eeq

\section{Examples}

In order to illustrate and test the above expressions for the one-loop integral supercoefficients,
we can compare these with known cases.

\subsection{Five-point NMHV superamplitude }

\begin{figure}[ht]
\begin{center}
\scalebox{0.70}{\includegraphics{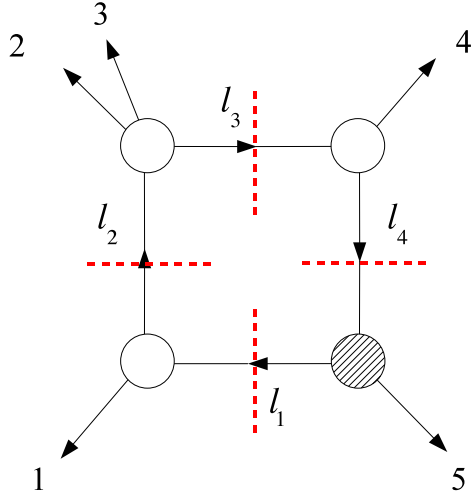}}
\end{center}
\caption{\it  A quadruple cut diagram determining the supercoefficient $\mathcal{C}(1,\{23\},4,5)$ in the five-point anti-MHV superamplitudes in
$\cN=4$ SYM and $\cN=8$ supergravity.
The black three-point amplitude has the anti-MHV configuration, and the remaining two amplitudes are MHV's. }
\label{antiMHV5quad}
\end{figure}

 The simplest case is the five-point NMHV superamplitude.
Here the relevant cut diagram we consider is depicted in  Figure  \ref{antiMHV5quad}.

The cut solution is:
\beq
\label{anti-MHVcutmom}
l_1 \, = \, \lambda_{l_1} \tilde\lambda_{1}\ , \qquad
l_2 \, = \, \lambda_{l_2} \tilde\lambda_{2}\ , \qquad
l_3 \, = \, \lambda_{l_3} \tilde\lambda_{3}\ , \qquad
l_4 \, = \, \lambda_{l_4} \tilde\lambda_{4}\ , \qquad
\eeq
and $\lambda_{l_1}, \ldots , \lambda_{l_4} $ are again easily determined
by imposing momentum conservation at the four corners of the cut diagram.
The result is
\beq
\label{anti-MHVcutmom2}
\lambda_{l_1} \, = \,  - { Q |4] \over [ 14]} \ , \qquad
\lambda_{l_2} \, = \,  { Q|1] \over [41]} \ , \qquad
\lambda_{l_3} \, = \,  - { P|1]  \over [41]} \ , \qquad
\lambda_{l_4} \, = \,   { P|4]  \over   [14]} \ , \qquad
\eeq
which is valid for a generic two-mass easy configuration, i.e.~for $P$ and $Q$ non null.
In the specific case of Figure \ref{antiMHV5quad}, we have $P=p_2 + p_3$ and $Q = p_5$.

It is instructive to first consider the $\mathcal{N}=4$ SYM
calculation in this case as the manipulations are very similar. Here the amplitude supercoefficient is given by the following fermionic integral,
\beqa
\label{antiMHV5pt}
\cC^{\cN=4} ( 1, \{ 2,3 \}, 4, 5 ) &\!\!\!\!=\!\!\!\!& {1\over 2} \sum_{\cS_{\pm}}\int\!\prod_{i=1}^4 d^4\eta_{l_i} \,
A^\mathrm{MHV}_{3} ( -l_1, 1, l_2)
\,
A^\mathrm{MHV}_{4} ( -l_2, 2, 3,  l_3)
\nonumber \\
&&\qquad\qquad\qquad
\times\ A^\mathrm{MHV}_{3} ( -l_3, 4, l_4)
\,
A^\mathrm{\overline{MHV}}_{3} ( -l_4, 5, l_1)
\ ,
\eeqa
where \cite{bhtrec,cahk}
\beqa
A^\mathrm{\overline{MHV}}_{3} ( -l_4, 5, l_1)& =& { \delta^{(4)} ( \eta_{l_4} [5 l_1] + \eta_5 [ l_1 l_4] + \eta_{l_1} [l_4 5 ] )\over
[l_4 5] [5 l_1] [l_1 l_4] }
\ ,
\eeqa
and the MHV superamplitudes are given by the usual formula \eqref{MHVNair}.
As in previous cases, there is only one solution to the cut equation for this amplitude, given in \eqref{anti-MHVcutmom} and
\eqref{anti-MHVcutmom2}.
Let us consider the fermionic integrations arising from \eqref{antiMHV5pt}. These give
\beqa
\int\!\prod_{i=1}^4 d^4\eta_{l_i}   \!\!\!\!\!\!\!\!&&
\delta^{(8)} (- \eta_{l_1} \lambda_{l_1} +  \eta_{l_2} \lambda_{l_2} + \eta_1 \lambda_1)
\, \delta^{(8)} (- \eta_{l_2} \lambda_{l_2} +  \eta_{l_3} \lambda_{l_3} + \eta_2 \lambda_2 + \eta_3 \lambda_3)
 \nonumber \\
 &&
\times\ \delta^{(8)} (- \eta_{l_3} \lambda_{l_3} +  \eta_{l_4} \lambda_{l_4} + \eta_4 \lambda_4)
 \, \delta^{(4)} ( \eta_{l_4} [5 l_1] + \eta_5 [ l_1 l_4] + \eta_{l_1} [l_4 5 ] )
\nonumber \\
= && \!\!\!\!\!\!\!  \left( {  \lan15\ran \lan23\ran[15] [45] \over [14]^2} \right)^4\,
\delta^{(8)} \Big( \sum_{i=1}^5 \eta_i \lambda_i\Big)  \,
\delta^{(4)} ( \eta_1 [23] + \eta_2 [31] + \eta_3 [12])
 \, ,
 \eeqa
where we have used the cut solution \eqref{anti-MHVcutmom} and \eqref{anti-MHVcutmom2}.

The contribution of the spinor factors arising from \eqref{antiMHV5pt} is readily evaluated to be
\beq
{[14]^8 \over \l23\r^4 ( \l15\r [15] \l45\r [45])^3 \bar{N}}
\ ,
\eeq
where $\bar{N} := [12][23][34][45][51]$.
Putting this together with the contribution from fermionic integrations we arrive at
the following result for the quadruple cut:
\beq
\cC^{\cN=4} ( 1, \{ 2,3 \}, 4, 5 )\ = \ {1\over 2} \, \delta^{(8)} \Big( \sum_{i=1}^5 \eta_i \lambda_i\Big)  \,
{\delta^{(4)} ( \eta_1 [23] + \eta_2 [31] + \eta_3 [12]) \over
\bar{N} \l 45 \r^4 } \, s_{15}\, s_{45}\ .
\eeq
We recall that  the five-point tree-level anti-MHV amplitude  is \cite{dhksgen}
\beq
A_{5}^{ \overline{\mathrm{MHV}}}(1,2,3,4,5)  \ = \  \delta^{(8)} \Big( \sum_{i=1}^5 \eta_i \lambda_i\Big)  \,
{\delta^{(4)} ( \eta_1 [23] + \eta_2 [31] + \eta_3 [12]) \over
\bar{N} \l 45 \r^4 }\ ,
\eeq
from which we conclude that the the supercoefficient is given by
\beq
\cC^{\cN=4}
( 1, \{ 2,3 \}, 4, 5 ) \ = \
 {s_{15}\, s_{45} \over 2} \, A_{5}^{ \overline{\mathrm{MHV}}}(1,2,3,4,5)
 \ ,
 \eeq
which is the expected result.

The expression for the supercoefficient $\cC^{\cN=8} ( 1, \{ 2,3 \}, 4, 5 ) $
in the  case of $ \cN=8$ supergravity
is again obtained by looking at the
quadruple cut depicted in Figure \ref{antiMHV5quad}, which in this case is
\beqa
\label{antiMHV5ptN=8}
\cC^{\cN=8}( 1, \{ 2,3 \}, 4, 5 ) &\!\!\!\!=\!\!\!\!&
{1\over 2} \sum_{\cS_{\pm}}\int\!\prod_{i=1}^4 d^8\eta_{l_i} \,
\cM^\mathrm{MHV}_{3} ( -l_1, 1, l_2)
\,
\cM^\mathrm{MHV}_{4} ( -l_2, 2, 3,  l_3)
\nonumber \\
&&\qquad\qquad\quad
\times\ \cM^\mathrm{MHV}_{3} ( -l_3, 4, l_4)
\,
\cM^\mathrm{\overline{MHV}}_{3} ( -l_4, 5, l_1)
\, ,
\eeqa
where
\beqa
\cM^\mathrm{MHV}_{3} ( -l_1, 1, l_2) &=& \big(A^\mathrm{MHV}_{3} ( -l_1, 1, l_2) \big)^2\ ,
\nonumber \\
\cM^\mathrm{\overline{MHV}}_{3} ( -l_4, 5, l_1)& =& \big(A^\mathrm{\overline{MHV}}_{3} ( -l_4, 5, l_1)\big)^2
\ ,
 \eeqa
and the four-point MHV superamplitudes is, using \eqref{eq:ElvangFreedmanMHV},
\beq
\label{fptp}
\cM^\mathrm{MHV}_{4} ( -l_2, 2, 3,  l_3)   = s_{-l_2 2} \big(A_\mathrm{4}^\mathrm{MHV}(-l_2, 2, 3, l_3) \big)^2 +
s_{-l_2 3} \big(A_\mathrm{4}^\mathrm{MHV}(-l_2, 3,2, l_3) \big)^2
\ .
\eeq
We can then recast $\cC^{\cN=8} ({1, \{ 2, 3 \}, 4, 5 })$ as  the sum of two terms,
\beq
\label{sumpl}
\cC^{\mathcal{N}=8} (1, \{2,3\}, 4, 5 )\ := \ \cC (1, \{2,3\},4, 5) \, + \,   \cC (1, \{3,2\}, 4, 5)
\ ,
\eeq
corresponding to the two terms in the sum in \eqref{fptp}. Each of these two terms is instantly obtained from the corresponding
result in Yang-Mills, we only have to calculate the dressing factors
\beqa
G(-l_2, 2, 3, l_3) & := &  s_{-l_2 2} \, = \, [12]{\l 23\r [34] \over [14]}  \,  = \, {\mathrm{Tr}_{+} ( 1234)   \over s_{14}}
\, ,
\nonumber \\
G(-l_2, 3, 2, l_3) & := &
s_{-l_2 3} \, = \, [13] {\l 32\r [24] \over [14]} \,  = \, {\mathrm{Tr}_{+} ( 1324)   \over s_{14}}
\  ,
\eeqa
using the explicit cut solution.

We can then write
\beq
\cC^{\cN=8} (1, \{2,3\},4, 5)  \ = \ {1\over 2}  \Big( s_{15}\, s_{45}  \, A_{5}^{\overline{\mathrm{MHV}}}(1,2,3,4,5) \Big)^2 \, {\mathrm{Tr}_{+} ( 1234)   \over s_{14}} \ ,
\eeq
or, for the coefficient of the pseudo-conformally invariant box function (obtained from the previous one by dividing by 
$-s_{15} s_{45}/2$),
\beq
\hat{\cC}^{\cN=8} (1, \{2,3\}, 4, 5 )  \ = \ - \Big( \, A_{5}^{ \overline{\mathrm{MHV}}}(1,2,3,4,5) \Big)^2 \,
\, \mathrm{Tr}_{+} ( 1234)\,  {s_{15} s_{45} \over s_{14}}    \ .
\eeq
A similar analysis applies for the coefficient of $ \cC^{\cN=8} (1, \{3,2\},4, 5)$.
We can now recast the supergravity coefficient \eqref{sumpl}   as a sum of squares of Yang-Mills coefficients, as
\beqa
\cC^{\mathcal{N}=8} (1, \{2,3\},4, 5) \ &=& 2\
\big(\cC^{\mathcal{N}=4}  (1, \{2,3\},4, 5)\big)^2 {\mathrm{Tr}_{+} (1234) \over s_{14}} 
\nonumber \\
 &+& 2\ \big(\cC^{\mathcal{N}=4}  (1,\{3,2\}, 4, 5 )\big)^2 {\mathrm{Tr}_{+} (1324) \over s_{14}}
\ .
\eeqa

We now wish to compare with the known results for the $\cN=8$ supergravity amplitude.
The sum in \eqref{sumpl} gives, modulo an overall common factor,
\beq
{1\over [12][34] } - {1\over [13][24]} \, = \, {[14][32]\over [12][13][24][34]}\ ,
\eeq
and we recognise that our present manipulations are the complex conjugate of those leading to \eqref{stide},
in particular
\beq
\sum_{\cP(23)} {G( -l_2, 2, 3, l_3) \over ([12][23][34])^2} \ = \ { \l 23\r\over [14] [23]^2 }   {[14][32]\over [12][13][24][34]}
\, = \,  {\l23\r\over [23]}{1\over [12][24][13][34]}
\ ,
\eeq
which is the complex conjugate of $ h(1,  \{2,3\}, 4)$.
Hence we can also write
\beqa
\nonumber
&& \cC^{\mathcal{N}=8} ( 1, \{ 2,3 \}, 4, 5 )
\ = \
\Big(
s_{15} s_{45}\,
\delta^{(8)} \Big( \sum_{i=1}^5 \eta_i \lambda_i\Big)  \,
 {
 \delta^{(4)} ( \eta_1 [23] + \eta_2 [31] + \eta_3 [12])\over [45][51] \l45\r^4}\Big)^2
 \nonumber \\
 \nonumber
 &&\qquad \qquad
  \times\  \sum_{\cP(23)} {G( -l_2, 2, 3, l_3) \over ([12][23][34])^2}
\\ \nonumber
   &&\qquad \qquad
   = \
\Big[ s_{15} s_{45}\,
\delta^{(8)} \Big( \sum_{i=1}^5 \eta_i \lambda_i\Big)  \,
 {\delta^{(4)} ( \eta_1 [23] + \eta_2 [31] + \eta_3 [12])\over  \l45\r^4}\Big]^2 \
 \nonumber \\ &&\qquad \qquad
 \times\ \ \  \bar{h} (1, \{2,3\}, 4) \, \bar{h} (4, \{5\}, 1)
 \ ,
 \label{fivefin}
 \eeqa
where we have also introduced $\bar{h} (4, \{5\}, 1) = 1  / ([45][51])^2$. Equation \eqref{fivefin} is in agreement with the results of
\cite{Bern:1998sv}.


\subsection{Six-point NMHV superamplitude}
\label{6ptNMHV}
In this section we will consider \eqref{eq:2mhend} in the case of six-point NMHV superamplitudes, and perform some numerical checks comparing our results to those derived  in  \cite{Bern:2005bb,BjerrumBohr:2005xx} for six-point NMHV graviton scattering amplitudes.
Specifically, we will compare our results to the following coefficients derived in  \cite{Bern:2005bb,BjerrumBohr:2005xx}: 
\beqa
\label{eq:6ptNMHVbernresult1}
  &&
  \mathcal{C}_{\rm 2mh}^{\mathcal{N}=8}(1^+,2^-,\{3^-,4^-\},\{5^+,6^+\})
 \,  =\,
  \\
  &&\quad
  \frac{1}{2}\,
  \frac{s_{34}s_{56}s_{12}^2 (x_{25}^2)^8}
       {[23][34][24][43]
        \langle 56 \rangle\langle 61 \rangle
        \langle 65 \rangle\langle 51\rangle
        [2 |x_{25} | 5 \rangle[2 |x_{25} |6 \rangle
        [3 |x_{25} | 1 \rangle[4 |x_{25} |1 \rangle}\  ,
\nonumber
\eeqa
and
\beqa
\label{eq:6ptNMHVbernresult2}
  &&
  \mathcal{C}_{\rm 2mh}^{\mathcal{N}=8}(3^+,4^-,\{5^+,6^+\},\{1^-,2^-\})
 \,  =\,
\\
  &&\quad
  \frac{1}{2}\,
  \frac{([3 | x_{14} | 4 \rangle)^8 s_{12} s_{56} (s_{34})^2}
       {\langle 45 \rangle\langle 46 \rangle
        \langle 56 \rangle\langle 65 \rangle
        [ 12 ] [ 13 ] [ 21 ] [ 23 ]
        [ 1 | x_{14}| 4 \rangle[ 2 | x_{14}| 4\rangle
        [ 3 | x_{14}| 5 \rangle[ 3 | x_{14}| 6\rangle}
  \nonumber\\
  &&\quad
  +\,
  \frac{1}{2}
  \frac{\langle 12 \rangle^6 [56]^6 s_{12}s_{56}s_{34}^2}
       {\langle 13 \rangle\langle 23 \rangle
        [ 45 ] [ 46 ]
        [ 4 | x_{14}| 1 \rangle[ 4 | x_{14}|2\rangle
        [ 5 | x_{14}| 3 \rangle[ 6 | x_{14}| 3 \rangle}\  .
        \nonumber
\eeqa

At six points, \eqref{eq:2mhend} is 
\beq
\mathcal{C}^{\cN=8}_{\rm 2mh}(i,i+1,P,Q)\ = \
\mathcal{C}^{\cN=8}_{\rm 3m}(i,i+1,P,Q)  \, +\,
\mathcal{C}^{\cN=8}_{\rm 3m}( i+1 , P, Q, i)
\ ,
\eeq
where $P=p_{i+2}+ p_{i+3}$ and $Q=p_{i+4}+ p_{i+5}$.
The three-mass supercoefficients given in \eqref{eq:3mend} become in this case
\beqa\label{eq:6pt3mone}
&&\mathcal{C}^{\cN=8}_{\rm 3m}(i,i+1,P,Q)\, =\,
  \sum_{{\mathcal{P}}(P)} \sum_{{\mathcal{P}}(Q)}
   \left(\mathcal{C}^{\cN=4}_{\rm 3m}(i,i+1,P,Q)\right)^2
  \\
   &&\qquad
   \times \ 2 \
   \frac{\langle i+2|P\, Q|i\rangle \langle i |i+1|i+2]}{\langle i |(i+1)\,  Q|i\rangle}\;
   \frac{\langle i+4|P\, (i+1)|i\rangle \langle i |Q|i+4]}{\langle i |(i+1)\, Q|i\rangle}
   \ ,
   \nonumber
\eeqa
and
\beqa\label{eq:6pt3mtwo}
&&\mathcal{C}^{\cN=8}_{\rm 3m}(i+1,P,Q,i)\, =\,
  \sum_{{\mathcal{P}}(P)} \sum_{{\mathcal{P}}(Q)}
   \left(\mathcal{C}^{\cN=4}_{\rm 3m}(i+1,P,Q,i)\right)^2
 \\
   &&\qquad
   \times\ 2\
   \frac{\langle (i+2)\,( i+1)\rangle \langle i+1 |i\, Q\, P | i+2]}{\langle i+1 |P\, i|i+1\rangle}\;
   \frac{\langle i+4|Q\, i|i+1\rangle \langle i+1 |P|i+4]}{\langle i+1 |P\, i |i+1\rangle}
   \ ,
\nonumber
\eeqa
where dressing factors involving one external leg are equal to one, and
the general expression for the $\mathcal{N}=4$ three-mass supercoefficient
entering \eqref{eq:6pt3mone} and \eqref{eq:6pt3mtwo} is given
in \eqref{eq:3mNequal4}.
Thus, we arrive at
\beqa
\label{newlabel}
&&\mathcal{C}^{\cN=8}_{\rm 2mh}(i,i+1,P,Q)\, =\,
  \sum_{{\mathcal{P}}(P)} \sum_{{\mathcal{P}}(Q)}
  \nonumber\\
   &&
  \bigg[
   \left(
   \frac{\delta^{(8)}\big(\sum_{i=1}^n \eta_i\lambda_i\big) R_{i;i+2\, i+4}   }{ \prod_j\l j j+1\r }
   \,
   \Delta_{i,i+1,i+2,i+4}\right)^2
   \nonumber\\
   &&
   \qquad
   \times \ 2 \
   \frac{\langle i+2|P\, Q|i\rangle \langle i |i+1|i+2]}{\langle i |(i+1)\, Q|i\rangle}\;
   \frac{\langle i+4|P\, (i+1)|i\rangle \langle i |Q|i+4]}{\langle i |(i+1)\, Q|i\rangle}
   \\
   &&
   +
   \left(
   \frac{\delta^{(8)}\big(\sum_{i=1}^n \eta_i\lambda_i\big) R_{i+1;i+4\, i}   }{ \prod_j\l j j+1\r }
   \,
   \Delta_{i+1,i+2,i+4,i}\right)^2
   \nonumber\\
   &&\qquad
   \times\ 2\
   \frac{\langle (i+2) (i+1) \rangle \langle i+1 |i\, Q\, P | i+2]}{\langle i+1 |P\, i|i+1\rangle}\;
   \frac{\langle i+4|Q\,  i| i+1\rangle \langle i+1 |P|i+4]}{\langle i+1 |P\, i |i+1\rangle}
   \bigg] .
   \nonumber
\eeqa

In order to be able to extract the coefficients for graviton amplitudes, we need to analyse the $\eta$-dependence of the 
$R$-functions  in  \eqref{newlabel}. 
The dependence on the supermomenta of the external particles  is contained in  the product
$[\delta^{(4)}(\Xi_{r;st})\delta^{(8)}(q)]^2$. Since we are going to compare to NMHV graviton amplitudes, 
we will only need the coefficients of terms of the form $(\eta_i)^8(\eta_j)^8(\eta_k)^8$.

Consider now the helicity assignment for the  coefficient in  \eqref{eq:6ptNMHVbernresult1}.
In \eqref{newlabel}, we encounter the quantities
$\Xi_{i;i+2\, i+4}$ and $\Xi_{i+1;i+4\, i+6}$. 
Therefore, we  consider the expressions 
$\delta^{(8)}(\Xi_{1;35})\delta^{(16)}(q)$ and
$\delta^{(8)}(\Xi_{2;51})\delta^{(16)}(q)$.
%
From \eqref{Xifunction}, we have, setting $i=1$, 
\beq
\Xi_{1;35}\, =\,
  \langle 1| (5+6)\,  4| 3\rangle \eta_3
  +\langle 1| (5+6)\, 4| 4\rangle \eta_4
  +\langle 34 \rangle [34] \langle 15 \rangle \eta_5
  +\langle 34 \rangle [34] \langle 16 \rangle \eta_6
  \ , 
\eeq
and 
\beq
\Xi_{2;51} =  \langle 21\rangle\langle 56\rangle
\,   ([61]\eta_5 +[15]\eta_6  +[56]\eta_1)
\, . 
\eeq

In the expansion of
\beq
  \delta^{(8)}(\Xi_{1; 35})\,
  \delta^{(16)}(\sum_{i=1}^{6}\eta_i\lambda_i)
  \ ,
\eeq
we need to pick the coefficient of $ (\eta_2)^8(\eta_3)^8(\eta_4)^8$, which is: 
\beqa
 &&\:
( \langle 1|5+6|3+4|2\rangle \langle 43 \rangle)^8,
 \eeqa
and in the expansion of
\beq
  \delta^{(8)}([61]\eta_5 +[15]\eta_6  +[51]\eta_1)
  \,
  \delta^{(16)}(\sum_{i=1}^{6}\eta_i\lambda_i)
\ ,
\eeq
the coefficient of  $ (\eta_2)^8(\eta_3)^8(\eta_4)^8$ vanishes.  In performing the sum in \eqref{newlabel} one will also need to include permutations of the above quantities. 

Now we turn to  the coefficient in \eqref{eq:6ptNMHVbernresult2} and compare to \eqref{newlabel}. 
In considering \eqref{newlabel} for this helicity assignment, we  encounter the quantities
$\Xi_{3;51}$ and $\Xi_{4;13}$. These  can be simply obtained by permuting indices in the expressions 
for  $\Xi_{1;35}$ and $\Xi_{2;51}$ given above. The corresponding coefficients for $ (\eta_1)^8(\eta_2)^8(\eta_4)^8$ 
are: 
\beqa
 (\langle 12\rangle \langle 34 \rangle s_{56})^8 \ , 
\eeqa
from $\Xi_{3;51}$, and 
\beqa
    \langle 4 |1+2|3]^8\ , 
\eeqa
from $\Xi_{4;13}$.

Now  we compare  \eqref{eq:6ptNMHVbernresult1}  and  \eqref{eq:6ptNMHVbernresult2}  to the  expansions of  
 $\mathcal{C}_{\rm 2mh}^{\mathcal{N}=8}(1,2,\{3,4\},\{5,6\})$  and
$\mathcal{C}_{\rm 2mh}^{\mathcal{N}=8}(3,4,\{5,6\},\{1,2\})$ which one derives from \eqref{newlabel}. 
Summing over the appropriate permutations,  we get
\beqa
\label{1}
 &&
  \mathcal{C}_{\rm 2mh}^{\mathcal{N}=8}(1^+,2^-,\{3^-,4^-\},\{5^+,6^+\})
 \,  =\, 
 \frac{\l 34\r^3 [56] \l 1|(5+6) \, (3+4)|2\r^6(s_{12}s_{234})^2}
      {2\l 1 2\r^6 \l 56 \r^2 \l 1|5+ 6|2]^2(s_{34})^2}
 \nonumber\\
 &&
 \quad\times
 \left[
 \frac{1 }
       { \l 16 \r [ 2 3]  
       \l 5 |3+ 4| 2]  \l 1| 5+ 6| 4]}
+ \frac{1 
          }
      {  \l 51\r  [2 3] 
      \l 6| 3+4|2]  \l 1|5+6|4] }  
 \right.
 \nonumber\\
 &&
 \qquad
 \left.
+ \frac{1
       }
      {  \l 61\r 
      [ 24] \l 5| 3+4| 2] 
      \l 1|5+ 6| 3]} 
+ \frac{1}
    { \l 15\r
      [24] \l 6| 3+4| 2]
      \l 1|5+6| 3] }
\right],
\eeqa
and
\beqa
\label{2}
 && \!\!\!\!\!\!\!\!\!\!
  \mathcal{C}_{\rm 2mh}^{\mathcal{N}=8}(3^+,4^-,\{5^+,6^+\},\{1^-,2^-\})
 \,  =\, 
 -
 \frac{\l 34\r^4 [34]^2 (s_{456})^2}
      {2\l 3|(1+2)\, (5+6)| 4\r^2}
 \times
 \nonumber\\
 && \!\!\!\!\!\!\!\!\!\!
 \quad
 \left[
  \frac{\l 12\r^6[12]\l 56\r [56]^6 }
     { \l 3|1+ 2| 4]^2 }
 \times
 \right.
 \nonumber\\
  && \!\!\!\!\!\!\!\!\!\!
  \qquad
\left(
\frac{1  }
     {\l 23\r [45]
       \l 1|5+6| 4]  
        \l 3|1+ 2|6] 
       }
- \frac{ 1 }
   {\l 13\r  [45]
      \l 2|5+6| 4]   
      \l 3| 1+2|6] } 
\right.
 \nonumber\\
 && \!\!\!\!\!\!\!\!\!\!
 \qquad
\left.
- \frac{1}
    {\l 23\r  [46] 
       \l 1|5+6| 4]  
        \l 3|1+ 2| 5] }
+ \frac{ 1}
   {\l 13\r [46] 
      \l 2| 5+6| 4]  
       \l 3|1+2|5] }
\right)
 \nonumber\\
 &&
 +
 \frac{\l 12\r [56] \l 4| 1+ 2| 3]^6}
     {\l 56\r^2 [12]^2}
 \times
 \nonumber\\
 &&
\left(
 \frac{1}
    {\l 45\r [23]   
        \l 4|5+ 6|1]  \l 6|1+2|3]}
- \frac{1}
    {\l 45\r [13] 
       \l 4|5+6|2] \l 6|1+2|3]}
\right.
 \nonumber\\
 && 
\left.\left.
- \frac{1}
    {\l 46\r [23]   
       \l 4|5+6|1] \l 5|1+2| 3]} 
+ \frac{1}
    {\l 46\r [13]  
        \l 4|5+6|2] \l 5|1+2|3] }
\right)\right]\, .
\eeqa
We have checked numerically that \eqref{1} and \eqref{2} agree with \eqref{eq:6ptNMHVbernresult1}
and \eqref{eq:6ptNMHVbernresult2}, respectively.

\section{General supergravity amplitudes}

Let us now consider going beyond NMHV. For the $\mathrm{N^2MHV}$ amplitudes we seek degree 16 (for $\cN=4$ SYM) or degree 32  (for $\cN=8$ supergravity) contributions which
leads to the following possibilities for the four tree amplitudes entering into the quadruple cuts:
one can have four MHV amplitudes, leading to four-mass, three-mass and two-mass coefficients,
or two MHV amplitudes, one anti-MHV amplitude and one NMHV amplitude, leading to three-mass and two-mass hard
coefficients, or two NMHV and two anti-MHV amplitudes leading to two-mass easy coefficients, or finally one can have one MHV amplitude, two anti-MHV amplitudes and one  $\mathrm{N^2MHV}$ amplitude,
leading to the two-mass easy and one-mass coefficients.

\begin{figure}[ht]
\begin{center}
\scalebox{0.70}{\includegraphics{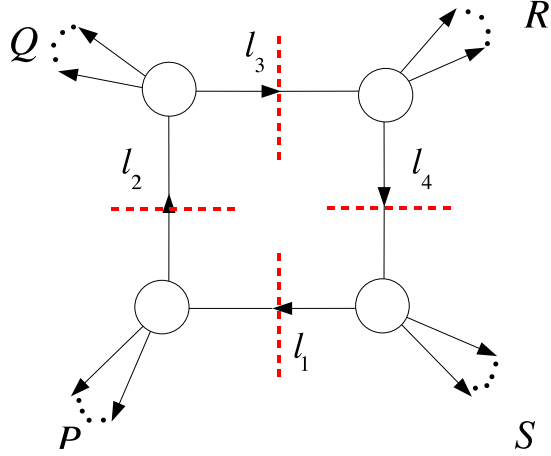}}
\end{center}
\caption{\it  A quadruple cut diagram determining the four-mass supercoefficient in an $N^2MHV$ amplitude. The four tree-level superamplitudes entering the cut have the MHV configuration.  }
\label{4mcoeff_fig}
\end{figure}

For the four-mass coefficients, the obvious quadruple cut diagram, represented in Figure \ref{4mcoeff_fig},  has four MHV tree-level superamplitudes, and is given by
\beqa\label{eq:4mstart}
&&\!\!\!\!\!\!\!\!\!\!\mathcal{C}^{\cN=8}_{\rm 4m}(P,Q,R,S)= \frac12 \sum_{\mathcal{S}_\pm}\int \prod_{i=1}^4 d^8\eta_{l_i}
\ {\cM}^{\rm MHV} (-l_1,P,l_2)\ \cM^\mathrm{MHV}(-l_2,Q,l_3) \nonumber\\
 &&\qquad\quad \qquad\quad \times\  \cM^\mathrm{MHV}(-l_3,R,l_4)\
\cM^\mathrm{MHV}(-l_4,S,l_1)\ .
\eeqa

Using \eqref{eq:ElvangFreedmanMHV} this is equal to
\beqa\label{eq:4mstart2}
&& \frac12 \sum_{\mathcal{S}_\pm}\int \prod_{i=1}^4 d^8\eta_{l_i} \sum_{\mathcal{P}(P,Q,R,S)}
\ [A^{\rm MHV}(-l_1,P,l_2)]^2\
G^{\rm MHV}(-l_1,P,l_2)
  \nonumber\\
  &&\quad
\times\
[A^{\rm MHV}(-l_2,Q,l_3)]^2\
G^{\rm MHV}(-l_2,Q,l_3)
\
[A^{\rm MHV}(-l_3,R,l_4)]^2\
G^{\rm MHV}(-l_3,R,l_4)
  \nonumber\\
  &&\quad
\times\
[A^{\rm MHV}(-l_4,S,l_1)]^2\
G^{\rm MHV}(-l_4,S,l_1)
\ ,
\eeqa
where the sum $\sum_{\mathcal{P}(P,Q,R,S)}$ is over permutations of momenta
within each of the sets of momenta $P$, $Q$, $R$ and $S$.
Since the dressing factors $G$ are independent of the superspace variables $\eta$, the
superspace integrals will only act on the square of the product of the four tree MHV superamplitudes
in the expression above. This is the same calculation as (the square of) the corresponding
$\mathcal{N}=4$ Yang-Mills four-mass coefficient, and hence one deduces that
\beqa\label{eq:4mend}
&&\!\!\!\!\!\!\!\!\!\!\mathcal{C}^{\cN=8}_{\rm 4m}(P,Q,R,S)= \frac12 \sum_{\mathcal{S}_\pm}
\sum_{\mathcal{P}(P,Q,R,S)}
\big(\mathcal{P}^{\rm 4m}_{n;1}\big)^2
G^{\rm MHV}(-l_1,P,l_2)
  \nonumber\\
  &&\quad
\times\
G^{\rm MHV}(-l_2,Q,l_3)
\
G^{\rm MHV}(-l_3,R,l_4)
\
G^{\rm MHV}(-l_4,S,l_1)
\ ,
\eeqa
where $\mathcal{P}^{\rm 4m}_{n;1}$ is the coefficient function given in equation
(5.11) of \cite{dhksgen} (this depends on the external momenta and in addition on the loop variables $l_i$, the solutions for which must be substituted).

A comment is in order here. We observe that, in contradistinction with the coefficients considered so far, because of the presence in \eqref{eq:4mend}  of a sum over the two solutions, we cannot recast immediately the right-hand side of this equation in terms of squares of $\cN=4$ supercoefficients; this appears to be  a general feature of four-mass box coefficients. 

\begin{figure}[ht]
\begin{center}
\scalebox{0.73}{\includegraphics{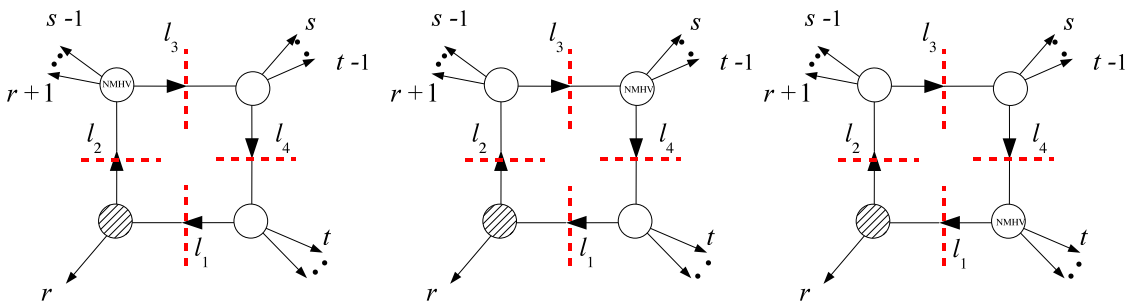}}
\end{center}
\caption{\it  Quadruple cut diagrams contributing to  the three-mass supercoefficient in an $N^2MHV$ amplitude.
Additional quadruple cut diagrams contributing to this supercoefficient are obtained as special cases of the
 four-mass quadruple cut diagram in Figure \ref{4mcoeff_fig}.  }
\label{3mcoeffNNMHV_fig}
\end{figure}

For the three-mass case, we have two possibilities. The first one corresponds to a special case of a four-mass coefficient, where  one of the four tree superamplitudes in Figure \ref{4mcoeff_fig} is a three-point MHV amplitude.
In addition, there  are  three new diagrams, represented in Figure \ref{3mcoeffNNMHV_fig}.
%

We focus our attention for instance on the second diagram in Figure \ref{3mcoeffNNMHV_fig}. This gives
\beqa\label{eq:3mstarter}
&&\!\!\!\!\!\!\!\!\!\!\mathcal{C}^{\cN=8}_{\rm 3m}(r,P,Q,R)\left.\right|_2= {1\over 2} \sum_{\mathcal{S}_\pm}\int \prod_{i=1}^4 d^8\eta_{l_i}
\ {\cM}_3^{\overline{\rm MHV}} (-l_1,r,l_2)\ \cM^\mathrm{MHV}(-l_2,P,l_3) \nonumber\\
 &&\qquad\quad \qquad\quad \times\  \cM^\mathrm{NMHV}(-l_3,Q,l_4)\
\cM^\mathrm{MHV}(-l_4,R,l_1)\ ,
\eeqa
where $P = \sum_{i=r+1}^{s-1}p_i$, $Q = \sum_{i=s}^{t-1}p_i$, $R = \sum_{i=t}^{r-1}p_i$.
Because of the presence of a three-point anti-MHV amplitude,
only one of the two cut solutions contributes to the cut diagram of  Figure \ref{MHVquad},
therefore one can then drop the sum over solutions in \eqref{eq:3mstarter}.
The explicit expressions  \eqref{eq:ElvangFreedmanMHV}, \eqref{boo} and \eqref{NMHV} may be inserted
into this relation, yielding
\beqa\label{eq:3mstarter2}
&&\!\!\!\!\!\!\!\!\!\!\mathcal{C}^{\cN=8}_{\rm 3m}(r,P,Q,R)\left.\right|_2 =
{1\over 2} \,  \int \prod_{i=1}^4 d^8\eta_{l_i} \sum_{\mathcal{P}(P,Q,R)}
\ [A^{\mathrm{\overline{MHV}}}_3(-l_1,r,l_2)]^2\
[A^{\rm MHV}(-l_2,P,l_3)]^2\
  \nonumber\\
  &&
\times\
G^{\rm MHV}(-l_2,P,l_3)
\
[A^{\rm MHV}(-l_3,Q,l_4)]^2\
\left(\sum\sum R^2(-l_3,Q,l_4) G^{\rm NMHV}(-l_3,Q,l_4)\right)
  \nonumber\\
  &&
\times\
[A^{\rm MHV}(-l_4,R,l_1)]^2\
G^{\rm MHV}(-l_4,R,l_1)
\ ,
\eeqa
where we indicate the NMHV summation schematically for simplicity. Again, the key point, as noted
in the discussion of NMHV amplitudes earlier, is that the fermionic variables corresponding to the loop momenta
do not appear in the dressing factors or the $R$-functions. Hence one can perform these superspace
integrations ignoring these functions -- and this corresponds to performing the same steps as in the
corresponding $\mathcal{N}=4$ Yang-Mills case, with the difference that the result is squared.
Thus we obtain
\beqa\label{eq:3mlast}
&&\!\!\!\!\!\!\!\!\!\!\mathcal{C}^{\cN=8}_{\rm 3m}(r,P,Q,R)\left.\right|_2 \, = \, 2
\sum_{\mathcal{P}(P,Q,R)}
\big(\mathcal{C}^{\cN=4}_{\rm 3m}(r,P,Q,R)\left.\right|_2 \,\big)^2
G^{\rm MHV}(-l_2,P,l_3)
  \nonumber\\
  &&\quad
\times\
\left( \sum\sum R^2(-l_3,Q,l_4) G^{\rm NMHV}(-l_3,Q,l_4)\right)
\
G^{\rm MHV}(-l_4,R,l_1)
\ ,
\eeqa
where by $\mathcal{C}^{\cN=4}_{\rm 3m}(r,P,Q,R)\left.\right|_2$ we mean the result of the same quadruple cut diagram evaluated for $\cN=4$ SYM.
The two-mass hard discussion goes along similar lines.

Finally, consider the two-mass easy case. 
There are four types of diagrams possible here. 
A first non-vanishing contribution is obtained as a  special case 
of the four-mass quadruple cut (see Figure \ref{4mcoeff_fig}), 
when two opposite corners of the diagram are three-point MHV amplitudes.

A second possibility is a special case of the three-mass contributions considered earlier in 
Figure \ref{3mcoeffNNMHV_fig}, where 
the MHV amplitude opposite to the anti-MHV three-point amplitude is also a three-point amplitude. 
This particular quadruple cut diagram will in general  vanish  as 
it would entail constraints on the external kinematics 
(this is not specific to the particular amplitudes considered here, 
but is a general feature of two-mass easy quadruple cuts where the two opposite 
three-point amplitudes cannot be one MHV and one anti-MHV).

\begin{figure}[ht]
\begin{center}
\scalebox{0.70}{\includegraphics{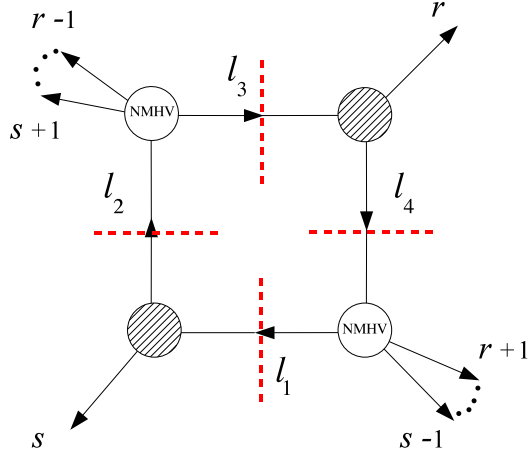}}
\end{center}
\caption{\it  Quadruple cut diagrams contributing to the two-mass easy coefficients of an
$N^2MHV$ amplitude. The black amplitudes have the anti-MHV helicity configuration.  }
\label{2mecoeff_diagsNNMHV}
\end{figure}

The third contribution comes from diagrams with two anti-MHV amplitudes at
opposite corners and two NMHV amplitudes at the other two corners, see Figure 
\ref{2mecoeff_diagsNNMHV}. This gives

\beqa\label{eq:2mecase3}
&&\!\!\!\!\!\!\!\!\!\!\mathcal{C}^{\cN=8}_{\rm 2me}(1,P,s,Q)\left.\right|_3= {1\over 2} \sum_{\mathcal{S}_\pm}\int \prod_{i=1}^4 d^8\eta_{l_i}
\ {\cM}_3^{\overline{\rm MHV}} (-l_1,1,l_2)\ \cM^\mathrm{NMHV}(-l_2,P,l_3) \nonumber\\
 &&\qquad\quad \qquad\quad \times\  {\cM}_3^{\overline{\rm MHV}}(-l_3,s,l_4)\
\cM^\mathrm{NMHV}(-l_4,Q,l_1)\ ,
\eeqa
where $P = \sum_{i=2}^{s-1}p_i$ and $Q = \sum_{i=s+1}^{n}p_i$.
Now we insert the expressions for the anti-MHV amplitudes and the NMHV
amplitudes, obtaining
\beqa\label{eq:2mecase3part2}
&&\!\!\!\!\!\!\!\!\!\!\mathcal{C}^{\cN=8}_{\rm 2me}(1,P,s,Q)\left.\right|_3= {1\over 2} \sum_{\mathcal{S}_\pm}\sum_{\mathcal{P}(P,Q)} \int \prod_{i=1}^4 d^8\eta_{l_i}
\Big( {A}_3^{\overline{\rm MHV}} (-l_1,1,l_2)\ A^\mathrm{MHV}(-l_2,P,l_3) \nonumber\\
 &&\qquad\quad \qquad\times\  A_3^{\overline{\rm MHV}}(-l_3,s,l_4)\
A^\mathrm{MHV}(-l_4,Q,l_1) \Big)^2
\left( \sum\sum R^2 G^{\rm NMHV}\right)(-l_2,P,l_3)
\nonumber\\
 &&\qquad\quad \qquad \times
\left( \sum\sum R^2 G^{\rm NMHV}\right)(-l_4,Q,l_1)
 \ ,
\eeqa
using a shorthand notation as previously. Only one solution to the loop
momenta conditions contributes, and one may perform the $\eta$ integrals directly - this is the same calculation as for the MHV two-mass
easy case, and thus we find the result
\beqa\label{eq:2mecase3part3}
&&\!\!\!\!\!\!\!\!\!\!\mathcal{C}^{\cN=8}_{\rm 2me}(1,P,s,Q)\left.\right|_3= {1\over 2} \sum_{\mathcal{P}(P,Q)}
\Big(\mathcal{C}^{\cN=4}_{\rm 2me}(1,P,s,Q)\Big)^2
\left( \sum\sum R^2 G^{\rm NMHV}\right)(-l_2,P,l_3)
\nonumber\\
 &&\qquad\quad \qquad\quad \times
\left( \sum\sum R^2 G^{\rm NMHV}\right)(-l_4,Q,l_1)
 \ ,
\eeqa
where the solutions for the loop momenta need to be inserted into the
terms containing the dressing functions $R$ and $G$.

\begin{figure}[ht]
\begin{center}
\scalebox{0.70}{\includegraphics{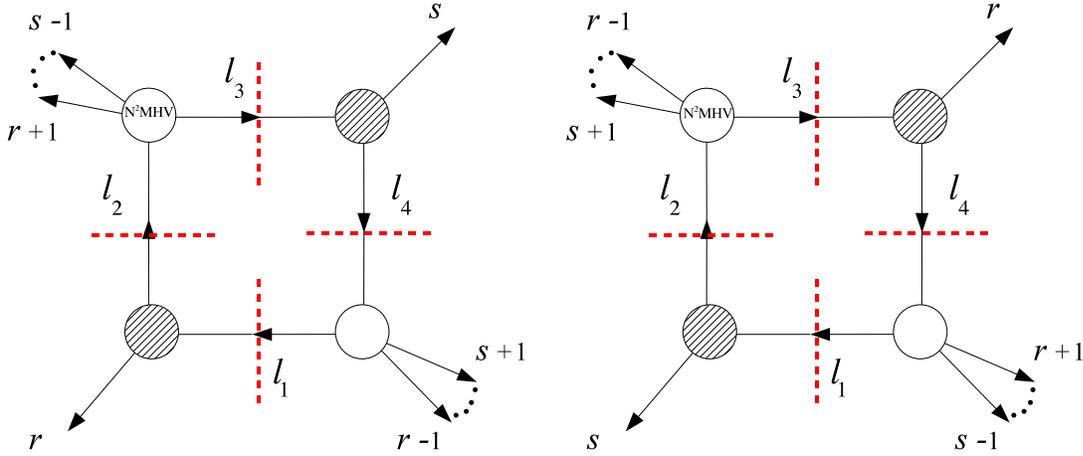}}
\end{center}
\caption{\it  Quadruple cut diagrams contributing to the two-mass easy coefficients of an
$N^2MHV$ amplitude.  }
\label{2mecoeff_diagsNNMHV_diagram1}
\end{figure}

Lastly, there is the two-mass easy diagram represented in Figure \ref{2mecoeff_diagsNNMHV_diagram1},
where  a new ingredient is the presence of a tree-level   $\mathrm{N^2MHV}$ amplitude.
This has been given in \cite{Drummond:2009ge}, and we reproduce this here:
\beqa\label{MNNMHV}
&&\!\!\!\!\!\!\!\!\!\!{\mathcal{M}}^{\rm N^2MHV}(1,\ldots,n)=
\sum_{{\mathcal{P}}(2,\ldots,n-1)}
[A^{\rm MHV}(1,\ldots,n)]^2
\nonumber\\
  &&\quad\times\ \sum_{2\leq a,b \leq n-1}
R_{n;a b}^2\Bigl[\sum_{a \leq c,d <
b} (R^{ba}_{n;ab;cd})^2 H^{(1)}_{n;ab;cd}
+ \sum_{b \leq c,d < n}
(R_{n;cd}^{ab})^2 H^{(2)}_{n;ab;cd} \Bigr]\, .
\eeqa

Explicit formulae for the $H$- and $R$-functions are given in \cite{Drummond:2009ge};
for our purposes we will only need to know the fact that the $H$ functions are independent of the superspace
variables $\eta$, and the $R(1,\ldots,n)$ functions do not depend on $\eta_1$ or $\eta_n$ -- the latter can
be seen from the fact that these extremal values are never taken by the subscripts on the $R$'s, and the explicit form they take (see (2.14) of \cite{Drummond:2009ge}).
Let us write the above equation in the short-hand form
\beqa\label{MNNMHVshort}
&&\!\!\!\!\!\!\!\!\!\!{\mathcal{M}}^{\rm N^2MHV}(1,\ldots,n)=
\sum_{{\mathcal{P}}(2,\ldots,n-1)}
[A^{\rm MHV}(1,\ldots,n)]^2\ \sum\sum R^2R^2H(1,\ldots,n)
\, .
\eeqa
Now we may write the quadruple cut for the two-mass easy diagrams as
\beqa\label{eq:2msetarter}
&&\!\!\!\!\!\!\!\!\!\!\mathcal{C}^{\cN=8}_{\rm 2me}(r,P,s,Q)\left.\right|_4= \frac12 \sum_{\mathcal{S}_\pm}\int \prod_{i=1}^4 d^8\eta_{l_i}
\ {\cM}_3^{\overline{\rm MHV}} (-l_1,r,l_2)\ \cM^\mathrm{N^2MHV}(-l_2,P,l_3) \nonumber\\
 &&\qquad\quad \qquad\quad \times\  {\cM}_3^{\overline{\rm MHV}}(-l_3,s,l_4)\
\cM^\mathrm{MHV}(-l_4,Q,l_1)\ .
\eeqa
As for the corresponding NMHV and MHV two-mass coefficients, only one of the two solutions to the cut condition contributes, given explicitly in \eqref{cutmom3}. Taking this into account, we get
\beqa
&&\!\!\!\!\!\!\!\!\!\!\mathcal{C}^{\cN=8}_{\rm 2me}(r,P,s,Q)\left.\right|_4 =
\frac12
\int \prod_{i=1}^4 d^8\eta_{l_i} \sum_{\mathcal{P}(P,Q)}
\ [A^{\rm \overline{MHV}}(-l_1,r,l_2)]^2\
[A^{\rm MHV}_3(-l_2,P,l_3)]^2   \nonumber\\
  &&\quad
\times\ \sum\sum R^2R^2H(-l_2,P,l_3)\
 [A^{\rm \overline{MHV}}(-l_3,s,l_4)]^2
  \nonumber\\
  &&\quad
\times\
[A^{\rm MHV}(-l_4,Q,l_1)]^2\
G^{\rm MHV}(-l_4,Q,l_1)
\ .
\eeqa
We may perform the loop superspace integrals and the final answer is
\beqa
&&\!\!\!\!\!\!\!\!\!\!\mathcal{C}^{\cN=8}_{\rm 2me}(r,P,s,Q) \left.\right|_4\, = \,
2  \sum_{\mathcal{P}(P,Q)}
(\mathcal{C}^{\cN=4}_{\rm 2me}(r,P,s,Q)\left.\right|_4)^2\
G^{\rm MHV}(-l_4,Q,l_1)  \nonumber\\
  &&\quad
\times\ \sum\sum R^2R^2H(-l_2,P,l_3)
\ ,
\eeqa
where the loop momenta are replaced by the cut solution in \eqref{cutmom3}.
For the one-mass case, the only contribution comes from the special case of the last two-mass easy case discussed immediately above -- that where $Q$ contains only one external momentum.

Having given some details of how the calculation proceeds for the $\mathrm{N^2MHV}$ case, one can see
how the general case will work. One can see from \cite{Drummond:2009ge} that the generalised $R$-functions
and dressing factors which arise in any quadruple cut do not depend upon the $\eta$ variables corresponding to
the loop momenta; hence one may perform the superspace integrals with these functions as spectators.
This calculation is however precisely the same as the corresponding $\mathcal{N}=4$ Yang-Mills case,
except that the coefficient is squared in the result. The outcome is that the $\mathcal{N}=8$ supergravity
coefficient is given by a sum of the squares of the result of the corresponding $\mathcal{N}=4$ Yang-Mills calculation,
factored into sums and products of $R$-functions and dressing factors. There is also in general a sum over solutions of the
cut equation, which need to be inserted into these expressions.
Thus we see how this approach yields $\mathcal{N}=8$ supergravity coefficients in terms of squares of
the results of $\mathcal{N}=4$ Yang-Mills calculations.

\section{Conclusions}

We have shown here in a number of cases how generalised
unitarity can be used in order  to generate new expressions for one-loop supercoefficients in 
$\mathcal{N}=8$ supergravity, and indicated how this applies in general.
In particular, using recent results for tree amplitudes \cite{Elvang:2007sg,Drummond:2009ge}, the one-loop supercoefficients take an intriguing form
involving sums of squares of $\mathcal{N}=4$ Yang-Mills one-loop expressions, times dressing
factors.  It seems likely that this structure will apply to all
one-loop supercoefficients in $\mathcal{N}=8$ supergravity.
It is certainly of interest to take this further, proving more general results in detail, deriving algorithms which produce the loop  dressing factors, and simplifying the expressions obtained when the solutions to the quadruple cut conditions for the 
loop  momenta are inserted. 
For the MHV case, it was easy to eliminate the loop momenta from the expressions we derived, and thus
find a direct correspondence with known results. It may be that a similar
outcome can be attained for non-MHV cases. The loop momenta solutions
are known explicitly, 
however the dressing factors entering non-MHV amplitudes are more complex.

It is intriguing that both tree-level superamplitudes and one-loop coefficients can be written in terms of squares of dual superconformal invariant quantities times bosonic dressing factors. It would be interesting to understand what possible deeper reasons may underly these regularities. In this context, we note that in \cite{dhks} it was shown that the dual superconformal invariant
$R$-functions appearing in the NMHV amplitudes in $\cN=4$ SYM have a coplanar twistor-space localisation.
It would be interesting if one could relate the simplicity of the tree-level and one-loop  results in $\cN=8$ supergravity
to simple twistor-space localisation properties. Interesting new ideas have been put forward recently
\cite{mask,ahetal,ho1} which in particular make a connection between on-shell recursion relations and twistor space \cite{ho2,mask,ahetal,ho1,wu}.

Underlying some of the work here are supersymmetric recursion relations \cite{bhtrec,cahk}. 
Interestingly,  there are additional recursion relations for  $\mathcal{N}=8$ supergravity amplitudes, 
arising from the fact that the tree amplitudes have a $1/z^2$ fall-off at large $z$ \cite{cahk,Spradlin:2008bu}. 
At present, the conditions imposed by  this constraint on one-loop amplitudes have not been much investigated. 
One might study the large-$z$ behaviour of the new expressions presented here for the supercoefficients and 
explore possible recursion relations for these, along the lines of \cite{Bern:2005hh}.

\section*{Acknowledgements}

It is a pleasure to thank Andi Brandhuber, Paul Heslop, Mark Spradlin,  Pierre Vanhove, Cristian Vergu and Anastasia Volovich
for interesting discussions.
GT would like to thank the Theory Group at Brown University for their warm hospitality and the opportunity to present the results 
of this paper.
This work was supported by the STFC under a Rolling Grant  ST/G000565/1.
GT is supported by an EPSRC Advanced Research Fellowship EP/C544242/1
and by an EPSRC Standard Research Grant EP/C544250/1.

\vspace{1cm}


\end{document}